\documentclass[a4paper,11pt]{article}
\pdfoutput=1
\usepackage{jheppub}
\usepackage[T1]{fontenc}
\usepackage{hyperref}
\usepackage{etoolbox}
    \makeatletter
    \patchcmd{\maketitle}{\@fpheader}{}{}{}
    \makeatother

\title{Quantum vacuum effects on the final fate of a collapsing ball of dust}

\author[a,b]{Hessamaddin Arfaei}
\author[a]{and Milad Noorikuhani}

\affiliation[a]{Department of Physics, Sharif University of Technology,\\P.O. Box 11155-9161, Tehran, Iran}
\affiliation[b]{School of Particles and Accelerators,\\Institute for Research in Fundamental Sciences(IPM),\\P.O. Box 19395-5531, Tehran, Iran}
\emailAdd{arfaei@ipm.ir}
\emailAdd{milad.nk70@gmail.com}

\abstract{We consider the quantum vacuum effects of the massless scalar fields that are non-minimally coupled to the background geometry of a collapsing homogeneous ball of dust. It is shown that for a definite range of coupling constants, there are repulsive quantum vacuum effects, capable of stopping the collapse process inside the black hole and precluding the formation of singularity. The final fate of the collapse will be a black hole with no singularity, inside which the matter stays balanced. The density of the final static matter will be close to the Planck density. We show that the largest possible radius of the stable static ball inside a black hole with Schwarzschild mass $M$ is given by ${{\left( \frac{1}{90\pi }\frac{M}{{{m}_{p}}} \right)}^{\frac{1}{3}}}{{\ell }_{p}}$. If the black hole undergoes Hawking radiation, the final state will be an extremal quantum-corrected black hole, with zero temperature, with a remnant of matter inside. We show that the resolution of singularity is not disrupted under Hawking radiation.}

\begin{document}
\maketitle
\flushbottom
\raggedbottom

\section{Introduction}
\label{sec:intro}
The final stage of the gravitational collapse is associated with the long-lasting problem of the singularity formation. Although the resolution of singularity is the task relegated to quantum gravity theories, we show that quantum vacuum effects in a semi-classical context, in certain conditions, are capable of accomplishing this task before quantum gravity effects become dominant. To this end, we consider a collapsing ball of homogeneous dust with spherical symmetry and introduce the quantum-originated stress-energy tensor in the interior and exterior geometries of this collapsing ball. Then, we will be able to predict the final fate of the collapse.

Classical models for the collapse of different configurations of matter~\cite{a,b,c,d} usually affirm the formation of singularity and, in addition, they can investigate its structure and visibility. However, to obtain a comprehensive model for the final fate of the collapse and the formation of singularity, one needs to use theories that include quantum effects. These theories can be based on quantum gravity or semi-classical models. One of the pivotal predictions of these models is the resolution of singularity problem. For instance, some articles have been devoted to the resolution of the singularity by arguments based on quantum gravity models~\cite{6,7,11,12,13,16,18}. In these models the quantum gravity effects will be important and dominant when the collapsing object approaches the limit that singularity --- associated with strong gravitational fields --- appears. These effects cause negative quantum-induced pressures that by producing repulsive forces can stop the collapsing object from continuing contraction. This process makes the final fate of the collapse devoid of singularity by preventing the collapsing object from crushing into zero physical volume. In other words, this prevents the density of matter and relevant scalars of general relativity from blowing up. The existence of negative pressures in the quantum gravity realm also would have some secondary outcomes for some realistic matter configurations. An example of such outcomes is the radiation of the collapsing matter due to the presence of negative pressures~\cite{radi}. This causes the shrinkage of the collapsing object due to the radiation of high energy matter fluxes. Such effects are claimed to result from the formation of naked singularities~\cite{80}. These secondary effects and similar ones are considered to provide probable empirical evidences for the indirect proof of some quantum gravity theories~\cite{a}. Quantum gravitational effects can also affect the formation of black hole horizons and Hawking radiation. In~\cite{bonanno} and~\cite{Hossenfelder}, interesting calculations of quantum gravity effects on the singularity resolution and the formation of horizon(s) can be found.

Another way of investigating the final fate of the gravitational collapse is using a model based on semi-classical approaches~\cite{casadio,barcelo}. One example of such approaches is quantum field theory in curved space-time. The reason for importance of such models is that the regions in the vicinity of the singularity are associated with strong gravitational fields within a very restricted volume. These conditions make the (geometry-dependant) quantum field theory effects an influential factor on the dynamics of the space-time. Therefore these regions are among the best places for the realization of the predictions of semi-classical theories. In this paper we use a semi-classical approach to the final fate of a collapsing homogeneous ball of dust by taking into account the vacuum expectation value of the stress-energy tensor of quantum scalar fields that are arbitrarily coupled to the background geometry of the collapsing object. In~\cite{jah.arf}, it was shown that for the special case of conformally invariant fields of arbitrary spin, the quantum vacuum effects are capable of preventing a collapsing thin shell of matter from reaching the singularity. A bounce radius at which the collapse of the shell stops and the direction of the movement reverses was predicted. In this regard, the singularity never forms. Here, we will consider arbitrarily (not necessarily conformally) coupled scalar fields in the collapse of a homogeneous ball of dust.

Our aim is to show that semi-classical approaches like considering quantum vacuum effects of the background geometry, are capable of addressing the singularity problem and even the information loss paradox before the quantum gravity becomes completely dominant. In sections~\ref{sec:two}, we introduce the basic requirements for calculating quantum-originated stress-energy tensor and by calculating it in section~\ref{sec:three}, we will be able to calculate the quantum corrections of arbitrarily coupled fields on the exterior geometry of the collapsing object. This is done by solving Einstein's equations ($G_{\mu\nu} = \frac{8\pi G}{c^4}\left\langle T_{\mu\nu} \right\rangle$) for a spherically symmetric space-time. Next, we calculate the effects of the quantum energy density and pressure, together with the dust density(matter density) on the dynamics of the interior geometry of the collapsing ball. Then, we will be equipped with enough tools to smoothly match the interior and exterior geometries at the boundary to make predictions about the final fate of the collapse. Section~\ref{sec:four} is devoted to this aim. We show that the existence of the quantum effects in both interior and exterior space-times can create repulsive Casimir forces which are able to prevent the collapsing object from continuing contraction and force it to be trapped inside the quantum corrected black hole --- that is formed during the collapse process --- in a stable static situation. We show that the largest possible radius of the ultimate stable static ball of dust is given by ${{\left( \frac{1}{90\pi }\frac{M}{{{m}_{p}}} \right)}^{\frac{1}{3}}}{{\ell }_{p}}$. It is also shown that for a typical initial condition at the onset of the collapse, the density of the matter at the final stable static situation is close to the Planck density and its physical radius will be some orders of magnitude larger than the Planck length. In addition, we show that the curvature of the interior geometry will be proportional to $\frac{1}{{{\ell }_{p}}^{2}}$ where ${\ell }_{p}$ is the Planck length. It is also demonstrated that the stability of the ball of dust, at the ultimate phase, is attained for a definite range of couplings.

The final outcome of this model will be a black hole within which the homogeneous ball of dust stays static with a finite, rather than zero, physical radius, which precludes the formation of singularity. Thus in this simple model which we have passed over many intricacies of natural gravitational collapses, we show that quantum field theory effects in a semi-classical level, have the ability to solve the singularity problem. In section~\ref{sec:four} we also argue that the inclusion of Hawking radiation will have no effect on the prevention of singularity formation. We will argue that Hawking radiation shrinks the black hole until it reaches an extremal phase with zero temperature which results in the stopping of the radiation. In this limit, ultimately, a finite remnant of matter remains balanced inside the extremal black hole and therefore makes it singularity-free. In addition, as is discussed in the last section, the existence of non-zero amount of matter within the black hole after the radiation allows the preservation of a finite amount of information. Potentially it can alleviate the information loss paradox.

\section{Stress-energy tensor of arbitrarily coupled quantum scalar fields}
\label{sec:two}
Since our aim is to study the quantum vacuum effects of scalar fields on curved background geometries, we start with the Lagrangian of the scalar fields which are non-minimally coupled to the background by coupling to Ricci scalar($R(x)$)~\cite{dav,parker},
\begin{equation}
\label{eq:lag}
L=\frac{1}{2}{{\left[ -g(x) \right]}^{\frac{1}{2}}}\left\{ {{g}^{\mu \nu }}(x)\Phi (x){{,}_{\mu }}\Phi {{(x)}_{,\nu }}+\left[ {{m}^{2}}+\xi R(x) \right]{{\Phi }^{2}}(x) \right\}
\end{equation}
where $g_{\mu\nu}$ is the metric of the background geometry and ${x}^{\mu }$ signifies the coordinates of space-time. The equation of motion is,
\begin{equation}
\label{eq:EoM}
\left[ \square -{{m}^{2}}-\xi R(x) \right]\Phi (x)=0 \,.
\end{equation}
here we have $\square \Phi ={{g}^{\mu \nu }}{{\nabla }_{\mu }}{{\nabla }_{\nu }}\Phi $ with sign convention: $(-+++)$.

Lagrangian \eqref{eq:lag} is invariant under conformal transformations ${{g}_{\mu \nu }}\to {{\bar{g}}_{\mu \nu }}={{\Omega }^{2}}(x){{g}_{\mu \nu }}$ if we have: $\xi =\frac{1}{4}\left[ \frac{n-2}{n-1} \right]$ with $n$ the dimension of space-time. Solutions of \eqref{eq:EoM} construct an orthonormal set and can be canonically quantized with the generalization of the familiar method in the Minkowskian quantum field theory. The stress-energy tensor reads
\begin{align}
\label{eq:enmomten}
{{T}^{\mu \nu }} &= {{\partial }^{\mu }}\Phi {{\partial }^{\nu }}\Phi -\frac{1}{2}{{g}^{\mu \nu }}{{\partial }^{\rho }}\Phi {{\partial }_{\rho }}\Phi -\frac{1}{2}{{g}^{\mu \nu }}{{m}^{2}}{{\Phi }^{2}}+\xi \left({{R}^{\mu \nu }}-\frac{1}{2}{{g}^{\mu \nu }}R\right){{\Phi }^{2}}\nonumber \\
&\quad +\xi \left[{{g}^{\mu \nu }}\square ({{\Phi }^{2}})-{{\partial }^{\mu }}{{\partial }^{\nu }}({{\Phi }^{2}})\right]
\end{align}
which with the help of \eqref{eq:EoM}, satisfies: ${{\nabla }_{\mu }}{{T}^{\mu \nu }}=0$.
A useful quantity is the trace of stress-energy tensor. This trace is already calculated for the special case of conformally coupled fields ($\xi=\frac{1}{6}$) as trace anomaly~\cite{dav,parker}. Here we aim to calculate this for arbitrarily coupled massless scalar fields. Since we are interested in quantized scalar fields, by Taking the expectation value of the trace of \eqref{eq:enmomten} and using \eqref{eq:EoM}, we obtain
\begin{equation}
\label{eq:trace}
\left\langle T_{\mu }^{\mu } \right\rangle =\left( 3\xi -\frac{1}{2} \right)\square \left\langle {{\Phi }^{2}} \right\rangle -{{m}^{2}}\left\langle {{\Phi }^{2}} \right\rangle \,.
\end{equation}
To calculate the quadratic terms in \eqref{eq:trace}, one can use Green's functions like Feynman propagator,
\begin{equation}
\label{eq:green}
{{G}_{F}}=-i\left\langle  0 \right|T(\Phi (x)\Phi ({x}'))\left| 0 \right\rangle \,,
\end{equation}
satisfying the equation,
\begin{equation}
\label{eq:greq}
\left( \square -{{m}^{2}}-\xi R \right){{G}_{F}}(x,{x}')=-\delta (x,{x}') \,.
\end{equation}
To find a solution for this equation, we can expand both sides in terms of Riemann normal coordinates ${{y}^{\mu }}$ for the point ${{x}^{\mu}}$ with the origin at ${{{x}'}^{\mu }}$ (see for example~\cite{bunch}). Then by working in the momentum space, we can obtain an expansion that enables us to attain a relation for the Green's function in momentum space by equating the same orders of the powers of Riemann normal coordinates in the both sides of the equation. Finally we can find the Green's function in the coordinate space by integrating over all momenta. The final relation is~\cite{j,k}
\begin{equation}
\label{eq:exp}
{{G}_{F}}(x,{x}')={{\Delta }^{\frac{1}{2}}}(x,{x}'){{(4\pi )}^{-\frac{n}{2}}}\int\limits_{0}^{\infty }{du{{(u)}^{-\frac{n}{2}}}}\exp \left[ -{{m}^{2}}u+\left( \frac{\sigma }{2u} \right) \right]F(x,{x}';u)
\end{equation}
where $\sigma =\sigma (x,{x}')=\frac{1}{2}{{y}_{\alpha }}{{y}^{\alpha }}$ is the one-half of the square of the proper distance between $x$ and ${x}'$ and $\Delta (x,{x}')$ is the Van Vleck determinant which is defined as~\cite{n}
\begin{equation}
\Delta (x,{x}')=-\det \left[ -{{\partial }_{\mu }}{{\partial }_{{{\nu }'}}}\sigma (x,{x}') \right]{{\left[ g(x)g({x}') \right]}^{-\frac{1}{2}}}
\end{equation}
The Green's function contains an expansion through function $F(x,{x}';u)$ which has the form
\begin{equation}
\label{eq:Fex}
F(x,{x}';u)=\sum\limits_{j=0}^{\infty }{{{a}_{j}}(x,{x}'){{(u)}^{j}}}
\end{equation}
where $u$ has its origin in,
\begin{equation}
\frac{1}{{{k}^{2}}-{{m}^{2}}+i\varepsilon }=\int\limits_{0}^{\infty }{du{{e}^{-u({{k}^{2}}-{{m}^{2}}+i\varepsilon )}}}
\end{equation}
and ($\varepsilon \to 0$). In fact The equation \eqref{eq:exp} is the generalization of the effective action method~\cite{schwarzqft} in Minkowski quantum field theory to it's curved space-time counterpart. In this method $-iu$ is usually referred to as DeWitt-Schwinger proper time. In the flat limit where all geometric quantities are zero, ${{a}_{0}}(x,{x}')=1$ and all higher order terms in \eqref{eq:Fex} vanish. Therefore, in this limit, the corresponding term for Green's function in flat space-time is reproduced.
With the use of Fynmann propagator in \eqref{eq:green}, one can obtain the quadratic terms in \eqref{eq:trace} as
\begin{equation}
\label{eq:limit}
\left\langle {{\Phi }^{2}} \right\rangle =\left\langle  0 \right|\Phi (x)\Phi (x)\left| 0 \right\rangle =\underset{x\to {x}'}{\mathop{\lim }}\,(i{{G}_{F}}(x,{x}')) \,.
\end{equation}
According to the definition, the limit ($x\to {x}'$) is identical to ${{y}^{\mu }}\to 0$ and therefore $\sigma \to 0$. Furthermore in this limit
$\Delta (x,x)=1$. Using these facts, \eqref{eq:exp} and \eqref{eq:limit}, the quadratic term takes the form
\begin{equation}
\label{eq:finT}
\left\langle {{\Phi }^{2}} \right\rangle =\underset{\sigma \to 0}{\mathop{\lim }}\,\left[ -{{(4\pi )}^{-\frac{n}{2}}}\int\limits_{0}^{\infty }{du{{(u)}^{-\frac{n}{2}}}}\exp \left[ -{{m}^{2}}u+\left( \frac{\sigma }{2u} \right) \right]F(x,x;u) \right] \,.
\end{equation}
In this limit the first three coefficients in \eqref{eq:Fex} are derived as~\cite{dav,parker}
\begin{subequations}\label{eq:coef}
\begin{align}
{{a}_{0}} & = 1 \,, \\
{{a}_{1}} & = \left( \frac{1}{6}-\xi  \right)R \,, \\
{{a}_{2}} & = \frac{1}{180}{{R}_{\alpha \beta \gamma \delta }}{{R}^{\alpha \beta \gamma \delta }}-\frac{1}{180}{{R}^{\alpha \beta }}{{R}_{\alpha \beta }}-\frac{1}{6}\left( \frac{1}{5}-\xi  \right)\square R+\frac{1}{2}{{\left( \frac{1}{6}-\xi  \right)}^{2}}{{R}^{2}} \,.
\end{align}
\end{subequations}

The vacuum expectation value for the trace of the stress-energy tensor of arbitrarily coupled quantum scalar fields is obtained by inserting \eqref{eq:coef} into \eqref{eq:finT}, and using the result in \eqref{eq:trace}. In the next section we use this trace to calculate the stress-energy tensor for the exterior geometry of a collapsing homogeneous ball of dust.
\section{Quantum corrections on the geometry of a collapsing homogeneous dust}
\label{sec:three}

The classical model of the homogeneous dust collapse was first proposed by Oppenheimer and Snyder and independently by Datt in the late 1930s~\cite{OS,Datt}. Here we consider a similar model and modify it by quantum vacuum effects. In this model the interior geometry of a spherically symmetric collapsing homogeneous dust is described by a closed contracting FRW metric. FRW metrics are prevalently used in describing homogeneous and isotropic universes in cosmology. Since the internal geometry of a spherically symmetric collapsing homogeneous object has the similar properties, it can be described by the same metric. In fact, this point was proved in the early models of spherically symmetric homogeneous dust collapse and we can also use it in our model. What we will do is involving the quantum-induced energy density and pressure in the FRW background of the inside of collapsing object and investigating their effects on the dynamics of collapse. The exterior geometry which is spherically symmetric will be quantum corrected Schwarzschild space-time. It is derived by adding the stress-energy tensor of the quantum fields in the ordinary Schwarzschild background, then solving Einstein's equations again with this quantum stress-energy tensor. This helps us to obtain quantum corrections to the Schwarzschild geometry. This procedure is similar to an iteration method since the quantum corrections are of order $\hbar $ with higher order terms neglected.

\subsection{Quantum corrected exterior geometry}

First we derive the quantum corrected exterior metric and after that we will investigate the quantum effects in the interior geometry. Using \eqref{eq:trace}, \eqref{eq:finT} and \eqref{eq:coef}, one obtains a relation for the trace of the quantum stress-energy tensor, namely
\begin{align}
\label{eq:schone}
\left\langle T_{\mu }^{\mu } \right\rangle &= \left(3\xi -\frac{1}{2}\right)\left(-\frac{1}{16{{\pi }^{2}}}\right)\square (\int\limits_{0}^{\infty }{\frac{du}{{{u}^{2}}}{{e}^{-{{m}^{2}}u}}+\int\limits_{0}^{\infty }{\frac{du}{u}{{e}^{-{{m}^{2}}u}}\left(\xi -\frac{1}{6}\right)}}R+\nonumber \\
&\quad +\int\limits_{0}^{\infty }{du}{{e}^{-{{m}^{2}}u}}(\frac{1}{180}{{R}_{\alpha \beta \gamma \delta }}{{R}^{\alpha \beta \gamma \delta }}-\frac{1}{180}{{R}^{\alpha \beta }}{{R}_{\alpha \beta }}-\frac{1}{6}\left(\frac{1}{5}-\xi \right)\square R+\frac{1}{2}{{\left(\frac{1}{6}-\xi \right)}^{2}}{{R}^{2}})\nonumber \\ &\quad +...) +\frac{1}{16{{\pi }^{2}}}{{m}^{2}}(\int\limits_{0}^{\infty }{\frac{du}{{{u}^{2}}}{{e}^{-{{m}^{2}}u}}+\int\limits_{0}^{\infty }{\frac{du}{u}{{e}^{-{{m}^{2}}u}}\left(\xi -\frac{1}{6}\right)}}R+\int\limits_{0}^{\infty }{du}{{e}^{-{{m}^{2}}u}}(\nonumber \\
&\quad \frac{1}{180}{{R}_{\alpha \beta \gamma \delta }}{{R}^{\alpha \beta \gamma \delta }} -\frac{1}{180}{{R}^{\alpha \beta }}{{R}_{\alpha \beta }}-\frac{1}{6}\left(\frac{1}{5}-\xi \right)\square R+\frac{1}{2}{{\left(\frac{1}{6}-\xi \right)}^{2}}{{R}^{2}})+...)
\end{align}
where the terms that are shown by ($...$) are related to the higher order terms in the expansion of \eqref{eq:Fex} at the limit ($x\to {x}'$). These terms are finite in $u=0$ which means they are UV-finite since according to the previous definitions $-iu$ acts like a scale parameter (DeWitt-Schwinger proper time). In this work we are solely interested in massless scalar fields. Performing the change of variables ${{m}^{2}}u=\lambda $ in the integrals of \eqref{eq:schone} and taking $m\to 0$, all UV-finite higher order terms will have powers of ${{m}^{2}}$ in their denominators and thus are IR-divergent. Such divergencies do not concern our present considerations and we omit them. Since in the Schwarzschild space-time we have: $R={{R}_{\mu \nu }}=0$, \eqref{eq:schone} simplifies to
\begin{equation}
\label{eq:fintr}
\left\langle T_{\mu }^{\mu } \right\rangle =\frac{1}{2880{{\pi }^{2}}}{{R}_{\mu \nu \gamma \delta }}{{R}^{\mu \nu \gamma \delta }} \,.
\end{equation}
It is clear that the final result is independent of the coupling constant $\xi$. Therefore, there is no difference between conformal and non-conformal field contributions to the renormalized trace of the quantum stress-energy tensor in the special case of Schwarzschild space-time. This means that our trace for arbitrarily coupled fields will be the same as the trace anomaly in this special case. Using ${{R}_{\mu \nu \gamma \delta }}{{R}^{\mu \nu \gamma \delta }}=48\frac{{{M}^{2}}}{{{r}^{6}}}$ for the Schwarzschild metric and plugging it into \eqref{eq:fintr}, results in
\begin{equation}
\label{eq:schtr}
\left\langle T_{\mu }^{\mu } \right\rangle =\frac{1}{60{{\pi }^{2}}}\frac{{{M}^{2}}}{{{r}^{6}}} \,.
\end{equation}
Having the trace of the stress-energy tensor, we need extra conditions to derive the complete form of this tensor. Spherical symmetry imposes two conditions.
One is $\left\langle T_{\theta }^{\theta } \right\rangle =\left\langle T_{\varphi }^{\varphi } \right\rangle $ and the other is~\cite{radial}:
$\left\langle T_{0}^{0} \right\rangle =\left\langle T_{r}^{r} \right\rangle $. A third condition comes from the conservation of stress-energy tensor
(${{\nabla }_{\nu }}\left\langle {{T}^{\mu \nu }} \right\rangle =0$) which leads to
\begin{equation}
4f(\left\langle T_{r}^{r} \right\rangle -\left\langle T_{\theta }^{\theta } \right\rangle )+(\left\langle T_{r}^{r} \right\rangle -\left\langle T_{0}^{0} \right\rangle )r{f}'+2rf\frac{d}{dr}\left\langle T_{r}^{r} \right\rangle =0
\end{equation}
where a prime denotes the derivative with respect to $r$ and $f=1-\frac{2M}{r}$. Using above constraints together with \eqref{eq:schtr}, one can calculate all elements of the stress-energy tensor. The result is
\begin{equation}
\label{eq:mat}
\left\langle T_{\mu }^{\nu } \right\rangle =-\frac{1}{120{{\pi }^{2}}}\left( \begin{matrix}
   1 & 0 & 0 & 0  \\
   0 & 1 & 0 & 0  \\
   0 & 0 & -2 & 0  \\
   0 & 0 & 0 & -2  \\
\end{matrix} \right)\frac{{{M}^{2}}}{{{r}^{6}}} \,.
\end{equation}
We assume the general form of the metric to be
\begin{equation}
d{{s}^{2}}=-{{e}^{\nu (r,t)}}d{{t}^{2}}+{{e}^{\lambda (r,t)}}d{{r}^{2}}+{{r}^{2}}(d{{\theta }^{2}}+{{r}^{2}}{{\sin }^{2}}\theta d{{\varphi }^{2}}) \,.
\end{equation}
As we showed before, our result for the quantum stress-energy tensor of the exterior geometry is independent of the coupling constant, and therefore, by putting the stress-energy tensor \eqref{eq:mat} into Einstein's equations ($R_{\mu }^{\nu }-\frac{1}{2}g_{\mu }^{\nu }R=8\pi \left\langle T_{\mu }^{\nu } \right\rangle $), the results will be similar to the results in~\cite{jah.arf} for conformally coupled fields. The solutions of Einstein's equations are easily obtained as
\begin{equation}
\label{eq:einss}
{{e}^{\nu }}={{e}^{-\lambda }}=\left( 1-\frac{{{\ell }_{p}}}{{{m}_{p}}}\frac{2M}{r}+\frac{{{\ell }_{p}}^{4}}{45\pi {{m}_{p}}^{2}}\frac{{{M}^{2}}}{{{r}^{4}}} \right)
\end{equation}
where we have worked in standard units for the final solutions by plugging the Planck mass ${{m}_{p}}$ and the Planck length ${{\ell }_{p}}$ into relations. Using \eqref{eq:einss}, the quantum corrected metric for the exterior geometry reads
\begin{align}
\label{eq:extfinalll}
d{{s}^{2}} &= -\left( 1-\frac{{{\ell }_{p}}}{{{m}_{p}}}\frac{2M}{r}+\frac{{{\ell }_{p}}^{4}}{45\pi {{m}_{p}}^{2}}\frac{{{M}^{2}}}{{{r}^{4}}} \right){{c}^{2}}d{{t}^{2}}+{{\left( 1-\frac{{{\ell }_{p}}}{{{m}_{p}}}\frac{2M}{r}+\frac{{{\ell }_{p}}^{4}}{45\pi {{m}_{p}}^{2}}\frac{{{M}^{2}}}{{{r}^{4}}} \right)}^{-1}}d{{r}^{2}}\nonumber + \\
&\quad +{{r}^{2}}d{{\Omega }^{2}} \,,
\end{align}
in which: $d{{\Omega }^{2}}=d{{\theta }^{2}}+{{\sin }^{2}}\theta d{{\varphi }^{2}}$. It is significant that ${{g}_{00}}$ in \eqref{eq:extfinalll} has two roots, showing the existence of two horizons. One is the outer horizon which is quantum corrected Schwarzschild horizon
\begin{equation}
\label{eq:ohor}
{{r}_{+}}\simeq \frac{2M}{{{m}_{p}}}{{\ell }_{p}}-\frac{1}{360\pi }\frac{{{m}_{p}}}{M}{{\ell }_{p}}.
\end{equation}
Note that we have kept only the linear terms in $\hbar$. The other horizon which is an inner one, arises due to the quantum corrections and takes the form
\begin{equation}
\label{eq:ihor}
{{r}_{-}}\simeq \frac{{{\left( \frac{1}{90\pi }\frac{M}{{{m}_{p}}} \right)}^{\frac{1}{3}}}}{1-{{\left( \frac{1}{90\pi }\frac{M}{{{m}_{p}}} \right)}^{\frac{1}{3}}}\frac{{{m}_{p}}}{6M}}{{\ell }_{p}} \,.
\end{equation}
Since a ${{\left( \frac{{{m}_{p}}}{M} \right)}^{\frac{2}{3}}}$ appears in the denominator and $\frac{{{m}_{p}}}{M}\ll 1$ for ordinary black holes, one can expand the fraction. In this expansion, the first three terms are of order lower than or equal to $\hbar$ and higher order terms can be neglected. We will return to these results for the quantum corrected exterior geometry of a collapsing object in the next section.

\subsection{Quantum vacuum energy in the interior geometry}

In this part we concentrate on the interior geometry of the collapsing homogeneous dust described by a contracting FRW metric. Our procedure begins with finding the quantum energy density and pressure of the massless scalar fields in this background. To find the quantum-induced energy density, we should sum over energies of the modes that are solutions of the equation of motion \eqref{eq:EoM} in a closed FRW background.

A closed FRW background is described by the line element
\begin{equation}
\label{eq:frw}
d{{s}^{2}}=-{{c}^{2}}d{{\tau }^{2}}+\frac{{R}'(r,\tau )}{1-k{{r}^{2}}}d{{r}^{2}}+{{R}^{2}}(r,\tau )d{{\Omega }^{2}}.
\end{equation}
The physical radius of a shell with the proper time $\tau$ and the proper radius $r$ is defined by
\begin{equation}
\label{eq:phrad}
R(r,\tau )=a(\tau )r
\end{equation}
where $a(\tau )$ is the scale factor. $k$ is a positive parameter for a closed geometry which is defined by $k=\frac{1}{{{\Re }^{2}}}$. $\Re$ is the radius of the curvature of the spatial metric which is a 3-sphere in this case. The physical radius \eqref{eq:phrad}, can define a proper area for the shells with different labels of $r$ and $\tau$ as:
\begin{equation}
C=\int{{{g}_{\theta \theta }}{{g}_{\varphi \varphi }}}d\theta d\varphi =\int{{{R}^{2}}(r,\tau )\sin \theta d\theta d\varphi =}4\pi {{R}^{2}}(r,\tau ) \,.
\end{equation}
In this model, if any shell of the homogeneous ball of dust is crushed into the singularity, its physical radius and proper area will be zero. According to \eqref{eq:phrad}, this is equivalent to saying that singularity for different shells is reached when the scale factor becomes zero. In our model, the scale factor $a(\tau)$ is independent of the proper radius of different shells. This means that, if there is a singularity formation at the final fate of the collapse, all shells will become singular simultaneously. We discuss this later where we show that the singularity would never form due to the quantum vacuum effects.

Now for a reason that will be clear later, we pay our attention to a static closed FRW background where the scale factor is constant. In this background the Ricci scalar is $R=\frac{6k}{{{a}^{2}}}$. By performing the change of variables $\sqrt{k}r=\sin \chi $ and using \eqref{eq:phrad}, \eqref{eq:frw} becomes
\begin{equation}
\label{eq:24}
d{{s}^{2}}=-{{c}^{2}}d{{\tau }^{2}}+\frac{{{a}^{2}}(\tau )}{k}(d{{\chi }^{2}}+{{(\sin \chi )}^{2}}d{{\Omega }^{2}}) \,.
\end{equation}
Using \eqref{eq:EoM}, the equation of motion will be
\begin{equation}
\label{eq:26}
(\square -\xi \frac{6k}{{{a}^{2}}})\Phi =0 \,.
\end{equation}
This equation should be solved in the background described by \eqref{eq:24}. The general solution can be written as
\begin{equation}
\label{eq:phiphi}
\Phi =\sum\limits_{n,l,m}{{{a}_{n,l,m}}}{{U}_{n,l,m}}+{{a}^{\dagger }}_{n,l,m}{{U}^{*}}_{n,l,m}
\end{equation}
where ${{a}_{n,l,m}}$ and ${{a}^{\dagger }}_{n,l,m}$ are annihilation and creation operators respectively after quantization. Therefore, the vacuum state can be defined as: ${{a}_{n,l,m}}\left| 0 \right\rangle =0$, for any set of quantum numbers $(n,l,m)$. The modes ${{U}_{n,l,m}}$ are eigenstates of \eqref{eq:26}, which are calculated as~\cite{ford1976}
\begin{equation}
\label{eq:27}
{{U}_{nlm}}={{A}_{nlm}}{{x}^{l}}C_{n-1}^{l+1}(\sqrt{1-{{x}^{2}}}){{Y}_{lm}}(\theta ,\varphi ){{e}^{-i{{\omega }_{n}}t}},
\end{equation}
which $x={{\left( \frac{\sin \chi }{\sqrt{k}} \right)}^{l}}$, $C_{n-l}^{l+1}$ are Gegenbauer functions and ${{Y}_{lm}}(\theta ,\varphi )$ are spherical harmonics. The quantum number $n$ takes the values $n=0,1,2,...$ and the quantum number $l$ has the values $l=0,1,...,n$ for fixed $n$. The eigenfrequencies are found to be
\begin{equation}
\label{eq:28}
{{\omega }_{n}}=\frac{c}{a}\sqrt{k\left[ {{(n+1)}^{2}}+6\xi -1 \right]}.
\end{equation}
The coefficients in \eqref{eq:27} are calculated by requiring the othonormality relation
\begin{equation}
({{U}_{\alpha }},{{U}_{\beta }})=-i\int{{{U}_{\alpha }}}{{\overset{\scriptscriptstyle\leftrightarrow}{{\partial }^{0}}}}{{U}^{*}}_{\beta }\sqrt{h}{{d}^{3}}x={{\delta }_{\alpha \beta }}
\end{equation}
where $\alpha$ and $\beta$ are sets of quantum numbers $(n,l,m)$. After determining these coefficients, one can use \eqref{eq:enmomten} to find the quantum energy density $\rho =-\left\langle T_{0}^{0} \right\rangle $ by using \eqref{eq:phiphi} for $\Phi$. The final relation for the quantum-induced energy density is
\begin{equation}
\label{eq:29}
\rho =\frac{\hbar {{k}^{{}^{3}/{}_{2}}}}{2{{\pi }^{2}}{a}^{3}}\sum\limits_{n=1}^{\infty }{{{(n+1)}^{2}}\frac{{{\omega }_{n}}}{2}} \,.
\end{equation}
We can also calculate the pressure which is defined as $p=\left\langle T_{1}^{1} \right\rangle $. The final result is $p=\frac{1}{3}\rho $. This is reminiscent of the equation of state for radiation and implies that massless scalar fields act like radiation in the static closed FRW background.

\begin{figure}[tbp]
\centering
\includegraphics[scale=0.8]{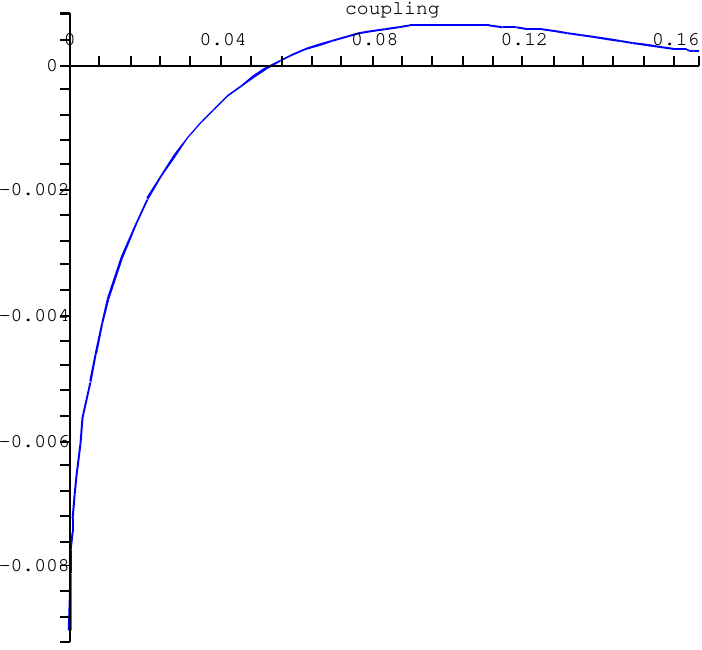}
\caption{\label{fig:coupli}A plot of $f(\xi)$ (vertical axis) as a function of coupling constant $\xi$~\cite{herdeiro}. There are negative values of $f(\xi)$ for a range of coupling constants which make the quantum-induced energy density \eqref{eq:31} and pressure \eqref{eq:neui} negative.}
\end{figure}

Considering \eqref{eq:29} and \eqref{eq:28}, the final relation for the quantum energy density reads
\begin{equation}
\label{eq:30}
\rho =\frac{\hbar {{k}^{2}}c}{4{{\pi }^{2}}{{a}^{4}}}\sum\limits_{n=0}^{\infty }{{{(n+1)}^{2}}\sqrt{{{(n+1)}^{2}}+6\xi -1}} \,.
\end{equation}
This expression for quantum-induced energy density is clearly divergent and one needs to apply renormalization techniques. For example one of the familiar methods is multiplying the terms of expansion in \eqref{eq:30} by a damping factor to get
\begin{equation}
\rho =\underset{\alpha \to 0}{\mathop{\lim }}\,\frac{\hbar {{k}^{2}}c}{4{{\pi }^{2}}{{a}^{4}}}\sum\limits_{n=0}^{\infty }{{{e}^{-\alpha n}}{{(n+1)}^{2}}\sqrt{{{(n+1)}^{2}}+6\xi -1}} \,.
\end{equation}
In this and other regularization techniques, the divergent part which can be regarded as the flat space contribution is subtracted. The remaining finite term is the result of the renormalization which in our case takes the form
\begin{equation}
\label{eq:31}
{{\rho }_{q}}=\frac{{{k}^{2}}\hbar c}{{{a}^{4}}}f(\xi ) \,.
\end{equation}
Hence for quantum pressure we have
\begin{equation}
\label{eq:neui}
{{p}_{q}}=\frac{1}{3}{{\rho }_{q}}=\frac{{{k}^{2}}\hbar c}{3{a}^{4}}f(\xi ) \,.
\end{equation}
Here $f(\xi )$ is a finite function of coupling constant which is derived after the renormalization. As an example we use the results in~\cite{herdeiro} for
the renormalization of \eqref{eq:30} which determine the values of $f(\xi)$. The result is shown in Figure~\ref{fig:coupli}. The important point is the existence of large negative values of $f(\xi )$ which make the quantum-induced energy density \eqref{eq:31} and pressure \eqref{eq:neui} negative. For the special case of the conformally invariant scalar fields with $\xi=\frac{1}{6}$, we have $f(\xi )=\frac{1}{480{{\pi }^{2}}}$ (see~\cite{dowker1977} for more detail) which implies a positive energy density. However, according to figure~\ref{fig:coupli}, for smaller values of the coupling constant the quantum energy density and pressure change sign. In the next section we will show that this negative values for the quantum vacuum energy density and pressure, together with the quantum corrections on the exterior geometry of the collapsing ball, cause important effects on the final fate of the collapse. Here we draw the reader's attention to the fact that negative energy densities and pressures are not due to negative kinetic energies of ghost fields but due to the quantum vacuum effects of massless scalar fields. In the next section we also show that the presence of matter density (${{\rho }_{m}}$) in the collapse process compensates for the negative values of the quantum vacuum energy density and pressure and makes the sums ${{\rho }_{m}}+{{\rho }_{q}}+{{p}_{q}}$ and ${{\rho }_{m}}+{{\rho }_{q}}$ positive. This implies that the weak energy conditions are satisfied.

\section{Quantum effects on the final fate of a homogeneous dust collapse}
\label{sec:four}

As we mentioned in the previous section, our scenario for the gravitational collapse of a homogeneous ball of dust with quantum vacuum effects is taking into account the dynamical effects of the quantum-induced energy density and pressure, in addition to the ordinary matter density. To this end, we consider a contracting closed FRW geometry as the interior space-time and a quantum corrected Schwarzschild geometry for the exterior space-time. Then we match the interior and exterior geometries to find out the final fate of the collapse. In this section, we follow this scenario and investigate the role of the quantum vacuum effects on the resolution of the black hole singularity.

\subsection{Quantum dynamical effects in the interior geometry}

The interior metric of collapsing homogeneous dust is described by \eqref{eq:frw} with $k>0$. The singularity for a shell is reached when its physical radius \eqref{eq:phrad} or basically the scale factor vanishes. Note that the central shell ($r=0$) is not regarded as a singular one while the scale factor is nonzero. In this background, the matter density of the collapsing dust is given by ${{\rho }_{m}}$, which from homogeneity condition takes the form
\begin{equation}
\label{eq:40}
{{\rho }_{m}}=\frac{{{\rho }_{0}}}{{{a}^{3}}} \,.
\end{equation}
In a closed FRW background the quantum energy density and the quantum pressure are given by
\begin{equation}
\label{eq:41}
{{\rho }_{q}}=\frac{{{k}^{2}}\hbar c}{{{a}^{4}}}f(\xi )+\Gamma ({{\dot{a}}^{4}},\ddddot{a},...)
\end{equation}
and
\begin{equation}
\label{eq:41prim}
{{p}_{q}}=\frac{{{k}^{2}}\hbar c}{3{{a}^{4}}}f(\xi )+\Pi ({{\dot{a}}^{4}},\ddddot{a},...) \,.
\end{equation}
The first terms in the right-hand side of both equations are the static space-time contributions \eqref{eq:31} and \eqref{eq:neui}. The second terms are contributions from the dynamical (contracting) space-time to the vacuum energy density and pressure of massless scalar fields. These terms can be derived from the calculation of $\left\langle T_{0}^{0} \right\rangle $ and $\left\langle T_{1}^{1} \right\rangle $ for a closed FRW space-time with a time-dependent (diminishing) scale factor. A dot denotes the derivative with respect to proper time $\tau$. $\Gamma$ and $\Pi$ involve terms of order $\frac{1}{{{\tau}^{4}}}$.

The Einstein's equations for a FRW metric leads to the familiar Friedmann equations. Using \eqref{eq:40} and \eqref{eq:41}, the first equation is
\begin{equation}
\label{eq:42}
{{\dot{a}}^{2}}-\gamma \frac{{{\rho }_{0}}}{a}-\gamma \left[ \frac{{{k}^{2}}\hbar }{{{a}^{2}}c}f(\xi )+\frac{{{a}^{2}}}{{c}^2}\Gamma ({{{\dot{a}}}^{4}},\ddddot{a},...) \right]=-k{{c}^{2}}\equiv E
\end{equation}
where we have defined $\gamma =\frac{8\pi G}{3}$ and $E$ is a constant. Using \eqref{eq:40}, \eqref{eq:41} and \eqref{eq:41prim}, the second Fridmann equation  takes the form
\begin{equation}
\label{eq:43}
\ddot{a}=-\frac{\gamma }{2}\frac{{{\rho }_{0}}}{{{a}^{2}}}-\frac{\gamma }{2}\left[ \frac{2{{k}^{2}}\hbar }{{{a}^{3}}c}f(\xi )+\frac{a}{{{c}^{2}}}\Gamma ({{{\dot{a}}}^{4}},\ddddot{a},...)+\frac{3a}{{{c}^{2}}}\Pi ({{{\dot{a}}}^{4}},\ddddot{a},...) \right] \,.
\end{equation}
With the help of these two equations, one can find the equations for the static scale factor $a={{a}_{0}}$ (the scale factor of a static background) where $\dot{a}$, $\ddot{a}$ and all other derivatives with respect to proper time are zero. The result is
\begin{subequations}\label{eq:y}
\begin{align}
\label{eq:y:44}
\gamma \frac{{{\rho }_{0}}}{{{a}_{0}}}+\gamma \frac{\hbar {{k}^{2}}}{{{a}_{0}}^{2}c}f(\xi ) & = k{{c}^{2}} \,,
\\
\label{eq:y:45}
\gamma \frac{{{\rho }_{0}}}{{{a}_{0}}^{2}}+2\gamma \frac{\hbar {{k}^{2}}}{{{a}_{0}}^{3}c}f(\xi ) & = 0 \,.
\end{align}
\end{subequations}
Now before evaluating ${a}_{0}$ we note that the equation \eqref{eq:42} is like a classical relation for the energy conservation in which the first term in the left-hand side acts like a kinetic term, the next two terms resemble the scale-dependent and the last term acts like a complicated velocity-dependent potential. The right-hand side is like the total amount of the conserved energy. First we look at the scale-dependent potential which is described by
\begin{equation}
\label{eq:46}
U(a)=-\gamma \frac{{{\rho }_{0}}}{a}-\gamma \frac{{{k}^{2}}\hbar }{{{a}^{2}}c}f(\xi )
\end{equation}
If we require the above potential to have a stationary point at $a={{a}_{0}}$, we should have: $\frac{\delta U(a)}{\delta a}=0$ at this point. Using \eqref{eq:46}, this condition implies
\begin{equation}
\gamma \frac{{{\rho }_{0}}}{{{a}_{0}}^{2}}+2\gamma \frac{\hbar {{k}^{2}}}{{{a}_{0}}^{3}c}f(\xi )=0
\end{equation}
which is the same \eqref{eq:y:45}. This means that the point $a={{a}_{0}}$, which is regarded as the scale factor of a static background, will be the stationary point of the scale-dependent potential regardless of its value. With the help of \eqref{eq:y:44} and \eqref{eq:y:45}, two unknown parameters can be determined. One more parameter should be chosen conventionally. We have two different choices. One is to choose a convention for $k$ and determine ${a}_{0}$ and ${\rho}_{0}$ and the other is to choose an arbitrary value for ${a}_{0}$ and determine $k$ and ${\rho}_{0}$. Both choices are physically equivalent. We follow the latter choice and set ${a}_{0}=1$ and determine $k$ and ${\rho}_{0}$ accordingly. Note that according to \eqref{eq:40}, for ${a}_{0}=1$ we have: ${{\rho }_{m}}={{\rho }_{0}}$, which means by determining ${\rho }_{0}$ we can find the density of the ball of dust in the stationary point. With the help of \eqref{eq:y:44} and \eqref{eq:y:45} we obtain,
\begin{equation}
\label{eq:48}
k=\gamma \frac{{{\rho }_{0}}}{2{{c}^{2}}} \,,
\qquad
{{\rho }_{m}}={{\rho }_{0}}=\frac{2{{c}^{5}}}{\left( -f(\xi ) \right)\hbar {{\gamma }^{2}}} \,.
\end{equation}
In closed FRW metrics we have: $k=\frac{1}{{{\Re }^{2}}}$ with $\Re$ the radius of curvature for the spherical spatial topology. Therefore by determining $k$ from \eqref{eq:48}, this radius can be calculated. It is clear from \eqref{eq:48} that to have a positive matter density in the static phase, there must be values of coupling constant for which $f(\xi )<0$. In figure~\ref{fig:coupli}, as an example for the calculation of $f(\xi )$, we can find many of such couplings. In fact from \eqref{eq:31} and \eqref{eq:neui}, the negative value of $f(\xi )$ for a specific coupling constant implies negative quantum-induced vacuum energy density and pressure. These negative quantities can produce repulsive vacuum effects that can stop the collapsing matter and cause a static phase. This can be easily seen in \eqref{eq:46} where the first term (matter contribution) acts like an attractive term of the potential and the next term (quantum vacuum contribution) acts like a repulsive one when $f(\xi )$ is negative. At the stationary point ($a={a}_{0}=1$), the collapsing dust enters a static phase which stems from a balance between the attractive and repulsive forces of the potential.

Now we investigate the condition for \eqref{eq:46} to be stable at the extremum point $a={a}_{0}=1$ i.e. $\frac{{{\delta }^{2}}U(a)}{\delta {{a}^{2}}}>0$. From \eqref{eq:46} and \eqref{eq:y:45} we get
\begin{equation}
{\left(\frac{{{\delta }^{2}}U(a)}{\delta {{a}^{2}}}\right)}_{a={a}_{0}=1}=-2\gamma {{\rho }_{0}}-6\gamma \frac{{{k}^{2}}\hbar }{c}f(\xi )=-2\gamma \frac{{{k}^{2}}\hbar }{c}f(\xi )>0 \,.
\end{equation}
This inequality is satisfied if we have
\begin{equation}
f(\xi )<0 \,.
\end{equation}
This condition implies that negative values of $f(\xi )$ which result in negative quantum densities and pressures, will also guarantee the stability of the stationary point $a={a}_{0}=1$. Henceforth, we only consider the coupling constants for which $f(\xi )$ is negative.

Remembering the definition $\gamma =\frac{8\pi G}{3}$ and plugging it into \eqref{eq:48} for the matter density in the stable static point we obtain
\begin{equation}
\label{eq:49}
{{\rho }_{m}}={{\rho }_{0}}=\frac{18}{{{\left( 8\pi  \right)}^{2}}\left( -f(\xi ) \right)}{{\rho }_{p}}
\end{equation}
where ${{\rho }_{p}}=\frac{{{m}_{p}}}{{{\ell }_{p}}^{3}}$ is the Planck density. This enormous density is consistent with our expectation, since at the stable static point the repulsive quantum contribution should be strong enough to cancel out the attractive matter contribution. This happens in such densities where strong quantum background effects are excited and cause an ultimate balanced phase.

\begin{figure}[tbp]
\centering
\includegraphics[scale=0.8]{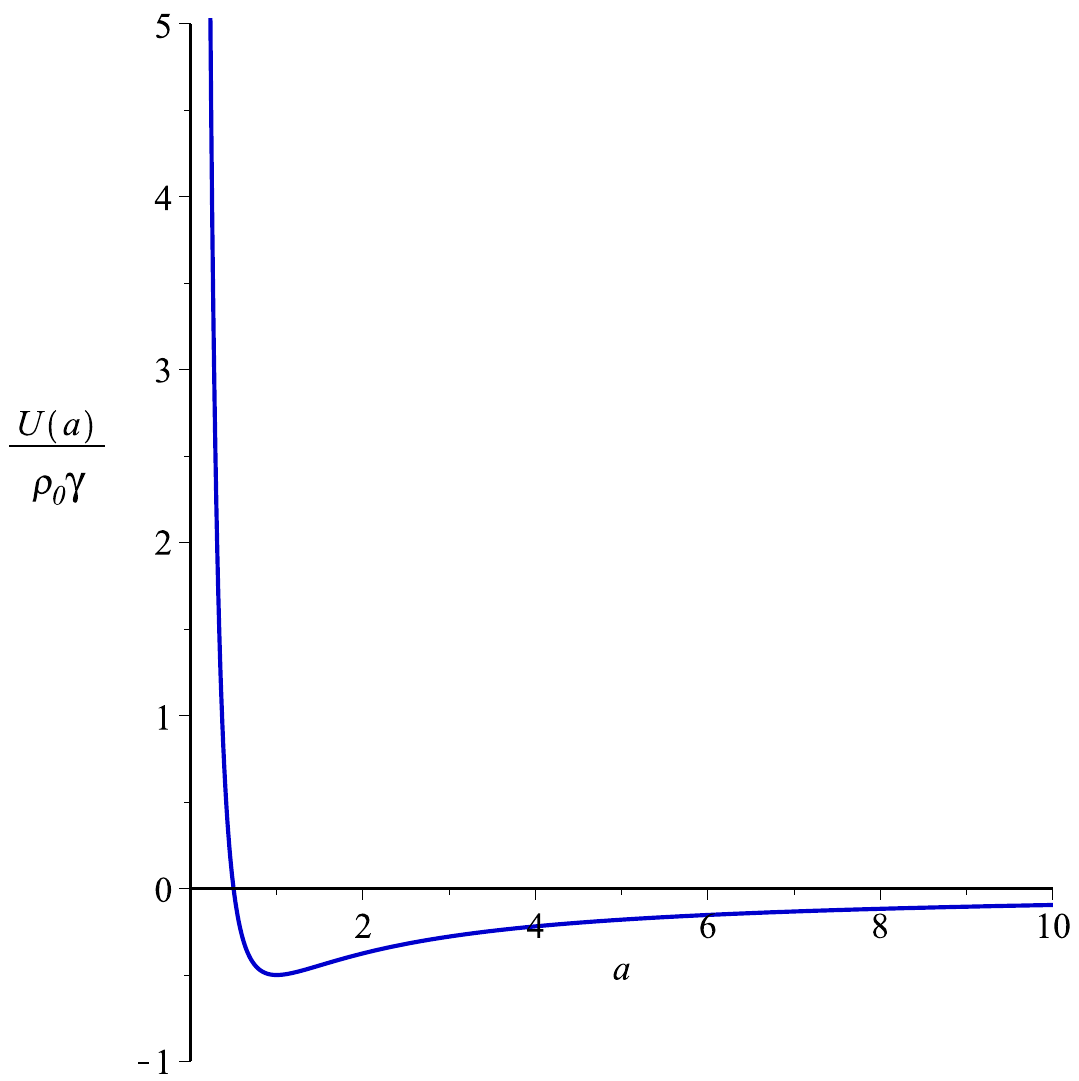}
\caption{\label{fig:pot}A plot of normalized potential $\frac{U(a)}{\gamma {{\rho }_{0}}}$ as a function of the scale factor. The point $a={{a}_{0}}=1$ is a stable stationary point for the potential. For smaller values of the scale factor the potential grows until it approaches infinity near $a=0$ which is the singularity point. The behavior of the normalized potential, which is reminiscent of a position-dependent potential in classical mechanics, implies that the collapse process should be stopped due to an increase in the potential of the system when the ball of dust continues contraction. Therefore the collapse process will be hindered due to the quantum-induced repulsive effects and the system will be settled at a stable static situation.}
\end{figure}

Regarding \eqref{eq:46} and using \eqref{eq:y:45} at ${a}_{0}=1$, we derive a useful relation for the scale-dependent potential that is
\begin{equation}
\label{eq:50}
U(a)=\gamma {{\rho }_{0}}(-\frac{1}{a}+\frac{1}{2{{a}^{2}}}) \,,
\end{equation}
where ${{\rho }_{0}}$ is given by \eqref{eq:49}. Figure~\ref{fig:pot} is a graph of normalized potential $\frac{U(a)}{\gamma {{\rho }_{0}}}$ as a function of the scale factor. As we anticipate, the static point $a={a}_{0}=1$ is an absolute minimum of the potential. For smaller values of the scale factor, the potential becomes larger until it reaches infinity at $a=0$ which is the singularity point. This behavior implies that after the collapsing matter has reached the stable static phase, it can not contract more towards a zero scale factor or equivalently the singularity, since it is not preferable for the system to increase its potential. Considering the equation \eqref{eq:42}, we argued that this equation is like a classical energy conservation relation with a kinetic term and a scale-dependent plus a complicated velocity-dependent potential. The latter potential is a complicated function of the time derivatives of the scale factor which are of order $\frac{1}{{{\tau}^{4}}}$. Furthermore, this potential is of order $\hbar$ because of the quantum nature of its origin, just like the quantum-originated repulsive term in the scale-dependent potential. This velocity dependent potential is clearly zero at the static point and is also very small at its proximity. therefore this term is negligible in such regions. Thus the behavior of the collapsing object is dictated by the scale-dependent potential near the static point.

Another important fact, which can be seen from \eqref{eq:50}, is the satisfaction of the weak energy conditions in the stable static phase despite the existence of the negative quantum densities and pressures. To see it, one should notice that in \eqref{eq:50}, the matter density comes with a negative sign ($-\frac{1}{a}$) and the quantum term comes with a positive sign ($+\frac{1}{2{{a}^{2}}}$). In the stable static phase we put $a={{a}_{0}}=1$ which leads to $U_{0}=\gamma {{\rho }_{0}}(-1+\frac{1}{2})$. This relation is a testament to the domination of the sign of the matter(dust) density in the stable static phase. Considering \eqref{eq:49}, we have ${\rho }_{m}={\rho }_{0}$ and therefore the relation for $U_{0}$ implies that the density of matter is twice as the quantum density (${{\rho}_{q}}=-\frac{1}{2}{{\rho }_{m}}$) in the static situation. Furthermore, according to \eqref{eq:neui} we have: ${{p}_{q}}=-\frac{1}{6}{{\rho }_{m}}$. Consequently we obtain
\begin{equation}
{{\rho }_{m}}+{{\rho }_{q}}=\frac{1}{2}{{\rho }_{m}}
\end{equation}
and
\begin{equation}
{{\rho }_{m}}+{{\rho }_{q}}+{{p}_{q}}=\frac{1}{3}{{\rho }_{m}}
\end{equation}
in the stable static phase. These two outcomes indicate that
\begin{equation}
{{\rho }_{tot}}\ge 0 \,,
\qquad
{{\rho }_{tot}}+{{p}_{tot}}\ge 0
\end{equation}
where ${\rho }_{tot}$ and ${p}_{tot}$ denote the total amount of density and pressure, available in the background, respectively. These relations stipulate that the energy conditions are satisfied. Therefore one should not worry about the occurrence of inconsistent physical situations in the stable static situation due to the existence of the negative quantum densities and pressures. Note that from \eqref{eq:mat}, considering $\rho =-\left\langle T_{0}^{0} \right\rangle $, $p=\left\langle T_{1}^{1} \right\rangle $, ${{p}_{\theta }}=\left\langle T_{2}^{2} \right\rangle $ and ${{p}_{\varphi }}=\left\langle T_{3}^{3} \right\rangle $, the same relations hold for the exterior geometry (for different components of pressure) and similarly the energy conditions are satisfied there.

As a summary, we have shown that quantum massless scalar fields with a definite range of couplings in the interior background, can create negative vacuum energy densities and pressures that cause repulsive effects. These effects can offset the attractive gravitational force of the collapsing matter and consequently hinder the collapse. This process can ultimately steer the collapsing object to a stable static phase with $a={{a}_{0}}=1$. In addition, we showed that the weak energy conditions are satisfied in this phase.

\subsection{Matching the geometries and final fate of the collapse}

In this part, by considering quantum vacuum effects, we match the exterior and interior geometries of a collapsing homogeneous ball of dust, using the results of previous parts. Then we investigate the physical properties of the stable static phase and discuss the final fate of the collapse.

The interior geometry which we label by ($-$), is described by \eqref{eq:frw}. The exterior geometry($+$), is described by \eqref{eq:extfinalll}, which can be rewritten as:
\begin{equation}
\label{eq:54}
d{{s}_{+}}^{2}=-f({{r}_{sc}}){{c}^{2}}d{{t}^{2}}+f{{({{r}_{sc}})}^{-1}}d{{r}_{sc}}^{2}+{{r}_{sc}}^{2}d{{\Omega }^{2}}
\end{equation}
where we have:
\begin{equation}
\label{eq:55}
f({{r}_{sc}})=1-\frac{2M}{{{r}_{sc}}}\frac{{{\ell }_{p}}}{{{m}_{p}}}+\frac{1}{45\pi }\frac{{{M}^{2}}{{\ell }_{p}}^{4}}{{{r}_{sc}}^{4}{{m}_{p}}^{2}} \,.
\end{equation}
Here ${{r}_{sc}}>{{R}_{c}}$ $\left( {{R}_{c}}\equiv R({{r}_{c}},\tau ) \right)$, which means the Schwarzschild radius is bigger than the physical radius of the boundary shell or equivalently the radius of the ball of dust.

We label the boundary surface between the interior and exterior regions by $\Sigma$. To follow the standard procedures of matching two geometries~\cite{poisson}, we should derive a relation for the metric of the boundary surface as seen from the either region. Firstly, we derive the line element of the boundary as seen from the interior region. By performing the change of variables $\sin \chi =\sqrt{k}r$, we obtain the line element \eqref{eq:24}. Now by considering this line element, the metric of the boundary surface as seen from the interior region can be obtained as
\begin{equation}
\label{eq:56}
d{{s}_{{{\Sigma }^{^{-}}}}}^{2}=-{{c}^{2}}d{{\tau }^{2}}+\frac{a{{(\tau )}^{2}}}{k}{{(\sin {{\chi }_{c}})}^{2}}d{{\Omega }^{2}} \,,
\end{equation}
where ${{\chi }_{c}}$ pertains to the boundary surface. To consider the boundary metric as seen from the exterior region, the schwarzschild time in the boundary $t$ should be related to the proper time $\tau$ of the boundary observers. In addition, the Schwarzschild radius ${r}_{sc}$ of the boundary should be related to the physical radius of the collapsing ball ${{R}_{c}}(\tau )$. These can be done by using the equations
\begin{equation}
\label{eq:57}
{{r}_{sc}}{{|}_{\Sigma }}={{R}_{c}}(\tau ) \,,
\qquad
t=T(\tau ) \,.
\end{equation}
In fact, the equation ${{r}_{sc}}{{|}_{\Sigma }}={{R}_{c}}(\tau )$ is the equation of motion for the boundary surface and $T(\tau )$ is a function that should be determined by the matching conditions. Considering \eqref{eq:54} and \eqref{eq:57}, the line element of the boundary as seen from the exterior region takes the form
\begin{equation}
\label{eq:58}
d{{s}_{{{\Sigma }^{^{+}}}}}^{2}=-{{c}^{2}}(F{{\dot{T}}^{2}}-\frac{{{F}^{-1}}{{{\dot{R}}}_{c}}^{2}}{{{c}^{2}}})d{{\tau }^{2}}+{{R}_{c}}^{2}d{{\Omega }^{2}}
\end{equation}
where a dot denotes the derivative with respect to proper time $\tau$. Considering \eqref{eq:55} and \eqref{eq:57}, $F$ is defined as
\begin{equation}
\label{eq:500}
F=f({{R}_{c}})=1-\frac{2M}{{{R}_{c}}}\frac{{{\ell }_{p}}}{{{m}_{p}}}+\frac{1}{45\pi }\frac{{{M}^{2}}{{\ell }_{p}}^{4}}{{{R}_{c}}^{4}{{m}_{p}}^{2}} \,.
\end{equation}

The first matching condition is the equality of the boundary metric as seen from the either region which implies $d{{s}^{2}}_{{{\Sigma }^{^{-}}}}=d{{s}^{2}}_{{{\Sigma }^{^{+}}}}$. Using \eqref{eq:56} and \eqref{eq:58}, this condition results in two useful equations
\begin{subequations}\label{eq:gip}
\begin{align}
\label{eq:gip:60}
{{R}_{c}}{{(\tau )}^{2}} & = \frac{a{{(\tau )}^{2}}}{k}{{(\sin {{\chi }_{c}})}^{2}} \,,
\\
\label{eq:gip:61}
F{{\dot{T}}^{2}}-\frac{{{F}^{-1}}{{{\dot{R}}}_{c}}^{2}}{{{c}^{2}}} & = 1.
\end{align}
\end{subequations}
The latter implies
\begin{equation}
\label{eq:defin}
F\dot{T}=\sqrt{\frac{{{{\dot{R}}}_{c}}^{2}}{{{c}^{2}}}+F}\equiv \beta \,.
\end{equation}

The second matching condition is the continuity of the extrinsic curvature which is defined as ${{K}_{ab}}={{n}_{\alpha ;\beta }}e_{a}^{\alpha }e_{b}^{\beta }$ where  ${{n}^{\mu }}$ is the unit normal vector to the boundary and $e_{\alpha }^{\beta }$ are the tangent basis vectors on the boundary surface. This condition can be shown by
\begin{equation}
\label{eq:62}
\left[ {{K}_{ab}} \right]={{K}_{ab}}^{+}-{{K}_{ab}}^{-}=0 \,.
\end{equation}
The unit normal vector to the boundary can be written as $n_{\mu }^{-}=(0,a,0,0)$ for the interior region and $n_{\mu }^{+}=(-\dot{R},\dot{T},0,0)$ for the exterior region. Again, a dot denotes the derivative with respect to proper time $\tau$. Using these facts, the nonvanishing components of the extrinsic curvature are~\cite{poisson},
\begin{equation}
\label{eq:63}
K_{\tau }^{\tau+ }=\frac{{\dot{\beta }}}{{\dot{{R}_{c}}}}\,,
\qquad
K_{\theta }^{\theta+ }=K_{\varphi }^{\varphi+ }=\frac{\beta }{{{R}_{c}}}
\end{equation}
for the exterior region and
\begin{equation}
\label{eq:64}
K_{\tau }^{\tau- }=0 \,,
\qquad
K_{\theta }^{\theta- }=K_{\varphi }^{\varphi- }=\frac{\cot {{\chi }_{c}}}{a(\tau )}\sqrt{k} \,.
\end{equation}
for the interior.

From $\left[ K_{\tau }^{\tau } \right]=0$, we conclude that $\beta$ is constant which, considering \eqref{eq:defin}, implies that $\frac{{{{\dot{R}}}_{c}}^{2}}{{{c}^{2}}}+F={{\beta }^{2}}$ is constant. This resembles an energy conservation relation for the collapsing ball. From $\left[ K_{\theta }^{\theta } \right]=\left[ K_{\varphi }^{\varphi } \right]=0$, by using \eqref{eq:defin} one obtains
\begin{equation}
\label{eq:65}
\sqrt{\frac{{{{\dot{R}}}_{c}}^{2}}{{{c}^{2}}}+F}={{R}_{c}}\frac{\cot {{\chi }_{c}}}{a(\tau )}\sqrt{k} \,.
\end{equation}
Using this equation and \eqref{eq:gip:60}, one derives
\begin{equation}
\label{eq:66}
\frac{{{{\dot{R}}}_{c}}^{2}}{{{c}^{2}}}+F={{(\cos {{\chi }_{c}})}^{2}} \,.
\end{equation}
Henceforth, we use the previous relations to obtain a remarkable equation which relates the Schwarzschild mass of the homogeneous ball of dust to its radius in the stable static phase. Starting from \eqref{eq:gip:60}, taking the time derivative of both sides yields
\begin{equation}
\label{eq:67}
k{{\dot{R}}_{c}}^{2}={{(\sin {{\chi }_{c}})}^{2}}{{\dot{a}}^{2}}.
\end{equation}
By substituting from \eqref{eq:42} for ${{\dot{a}}^{2}}$ in \eqref{eq:67} one obtains
\begin{equation}
\label{eq:68}
{{\dot{R}}_{c}}^{2}=\frac{{{(\sin {{\chi }_{c}})}^{2}}}{k}\left( -k{{c}^{2}}+\gamma \frac{{{\rho }_{0}}}{a}+\gamma \left[ \frac{{{k}^{2}}\hbar }{{{a}^{2}}c}f(\xi )+\frac{{{a}^{2}}}{{c}^2}\Gamma ({{{\dot{a}}}^{4}},\ddddot{a},...) \right] \right).
\end{equation}
As we discussed before, we can neglect the function $\left[\frac{{{a}^{2}}}{{c}^2}\Gamma ({{{\dot{a}}}^{4}},\ddddot{a},...) \right]$ since we are interested in exploring \eqref{eq:68} at (the proximity of) the static point. Using \eqref{eq:66} for ${{\dot{R}}_{c}}^{2}$ in \eqref{eq:68} one gets
\begin{equation}
{{c}^{2}}(1-{{(\sin {{\chi }_{c}})}^{2}}-F)=\frac{{{(\sin {{\chi }_{c}})}^{2}}}{k}(-k{{c}^{2}}+\gamma \frac{{{\rho }_{0}}}{a}+\gamma \frac{{{k}^{2}}\hbar }{{{a}^{2}}c}f(\xi )) \,,
\end{equation}
leading to
\begin{equation}
\label{eq:69}
{{c}^{2}}\left( 1-F \right)=\left( \gamma \frac{{{\rho }_{0}}}{a}\frac{{{(\sin {{\chi }_{c}})}^{2}}}{k}+\gamma \frac{k\hbar }{{{a}^{2}}c}f(\xi ){{(\sin {{\chi }_{c}})}^{2}} \right).
\end{equation}
Using \eqref{eq:gip:60}, one can substitute for ${{(\sin {{\chi }_{c}})}^{2}}$ in \eqref{eq:69} to obtain
\begin{equation}
\label{eq:70}
1-F=\gamma \frac{{{\rho }_{0}}}{{{c}^{2}}{{a}^{3}}}{{R}_{c}}^{2}+\gamma \frac{\hbar {{k}^{2}}{{R}_{c}}^{2}}{{{c}^{3}}{{a}^{4}}}f(\xi ).
\end{equation}
Here we set $a={{a}_{0}}=1$ which is the scale factor at the stable static situation, as we discussed before. Furthermore, by using \eqref{eq:48} for $k$, \eqref{eq:49} for ${{\rho }_{0}}$, \eqref{eq:500} for $F$ and by recalling the definition $\gamma =\frac{8\pi G}{3}$, the equation \eqref{eq:70}, after simplification, takes the form
\begin{equation}
\label{eq:71}
M-\frac{1}{90\pi }\frac{{{M}^{2}}}{{{R}_{c}}^{3}}\left( \frac{{{\ell }_{p}}^{3}}{{{m}_{p}}} \right)=\frac{4\pi }{3}{{R}_{c}}^{3}(\frac{9}{{{(8\pi )}^{2}}(-f(\xi ))}){{\rho }_{p}}.
\end{equation}
This equation is a relation between Schwarzschild mass and the physical radius of the homogeneous ball of dust at the stable static phase. Note that in this equation Schwarzschild mass includes two contributions. one is the ordinary matter (dust) contribution in the right hand side. Clearly, this term is the physical volume ($\frac{4\pi }{3}{{R}_{c}}^{3}$) times matter density at the stable static point. The other contribution is the second term in left hand side of \eqref{eq:71} which is a purely quantum-originated term, emanated from the vacuum effects of the interior and exterior geometries. Because of this new term, Schwarzschild mass is different from the ordinary mass ($M_{m}$).

For simplicity we can write
\begin{equation}
\label{eq:72}
{{R}_{c}}={{\gamma }_{r}}{{\ell }_{p}} \,,
\qquad
M={{\gamma }_{m}}{{m}_{p}} \,
\end{equation}
where we have introduced the parameters ${{\gamma }_{m}}$ and ${{\gamma }_{r}}$ to write the Schwarzschild mass and the physical radius of the static ball in terms of the Planck mass and the Planck length respectively. Using these new parameters, the equation \eqref{eq:71} can be rewritten as
\begin{equation}
\label{eq:73}
{{\gamma }_{m}}-\frac{1}{90\pi }\frac{{{\gamma }_{m}}^{2}}{{{\gamma }_{r}}^{3}}=\frac{3}{16\pi (-f(\xi ))}{{\gamma }_{r}}^{3} \,.
\end{equation}
This expression is a quadratic equation in terms of ${{\gamma }_{m}}$. The discriminant of the equation is
\begin{equation}
\label{eq:74}
\Delta =1+4\left( \frac{1}{90\pi } \right)\left( \frac{3}{16\pi f(\xi )} \right).
\end{equation}
To have positive real roots we require $\Delta \ge 0$, which leads to
\begin{equation}
\label{eq:501}
-f(\xi )\ge \frac{1}{120{{\pi }^{2}}} \,.
\end{equation}
For values of $f(\xi )$ satisfying this condition, we obtain real and positive solutions for ${\gamma }_{m}$,
\begin{equation}
\label{eq:75}
{{\gamma }_{m}}=\frac{-1\pm \sqrt{1+\frac{1}{120{{\pi }^{2}}(-f(\xi ))}}}{-\frac{2}{90\pi }}{{\gamma }_{r}}^{3}.
\end{equation}
With this equation, one can find a straightforward way to calculate the physical radius of the ball of dust at the ultimate static point for a given initial mass. Before setting a physical example for such calculation, we investigate the location of the stable static ball of dust. This helps us to understand whether the collapsing ball ultimately ends up in a black hole. To this end, we recall the quantum corrected horizons derived in previous section. Using \eqref{eq:ohor}, \eqref{eq:ihor} and \eqref{eq:72}, one finds the horizons as
\begin{subequations}\label{eq:hior}
\begin{align}
\label{eq:hior:76}
{{r}_{+}} & = 2{{\gamma }_{m}}{{\ell }_{p}}-\frac{1}{(360\pi ){{\gamma }_{m}}}{{\ell }_{p}} \,,
\\
\label{eq:hior:77}
{{r}_{-}} & = \frac{{{\left( \frac{{{\gamma }_{m}}}{90\pi } \right)}^{\frac{1}{3}}}}{1-{{\left( \frac{{{\gamma }_{m}}}{90\pi } \right)}^{\frac{1}{3}}}\frac{1}{6{{\gamma }_{m}}}}{{\ell }_{p}} \,.
\end{align}
\end{subequations}
Using \eqref{eq:75} and requiring the condition \eqref{eq:501}, the maximum value of the ${\gamma }_{r}$ is obtained as ${{\left( {{\gamma }_{r}} \right)}_{\max }}={{\left( \frac{{{\gamma }_{m}}}{90\pi } \right)}^{\frac{1}{3}}}$ in the limit $\left( -f(\xi ) \right)\longrightarrow \infty $. Therefore, considering \eqref{eq:72}, the largest possible value of the physical radius of the stable static ball is
\begin{equation}
\label{eq:78}
{{\left( {{R}_{c}} \right)}_{\max }}={{\left( \frac{{{\gamma }_{m}}}{90\pi } \right)}^{\frac{1}{3}}}{{\ell }_{p}}={{\left( \frac{1}{90\pi }\frac{M}{{{m}_{p}}} \right)}^{\frac{1}{3}}}{{\ell }_{p}} \,.
\end{equation}
Comparing this radius with the radii of the horizons in \eqref{eq:hior:76} and \eqref{eq:hior:77}, it is easy to see that the maximum possible value of the physical radius is always smaller than the radius of the inner horizon. This interesting result implies that the final fate of the collapse will always be a black hole inside which a stable static ball of dust is trapped with a radius smaller than the inner horizon.

\begin{figure}[tbp]
\centering
\includegraphics[scale=0.8]{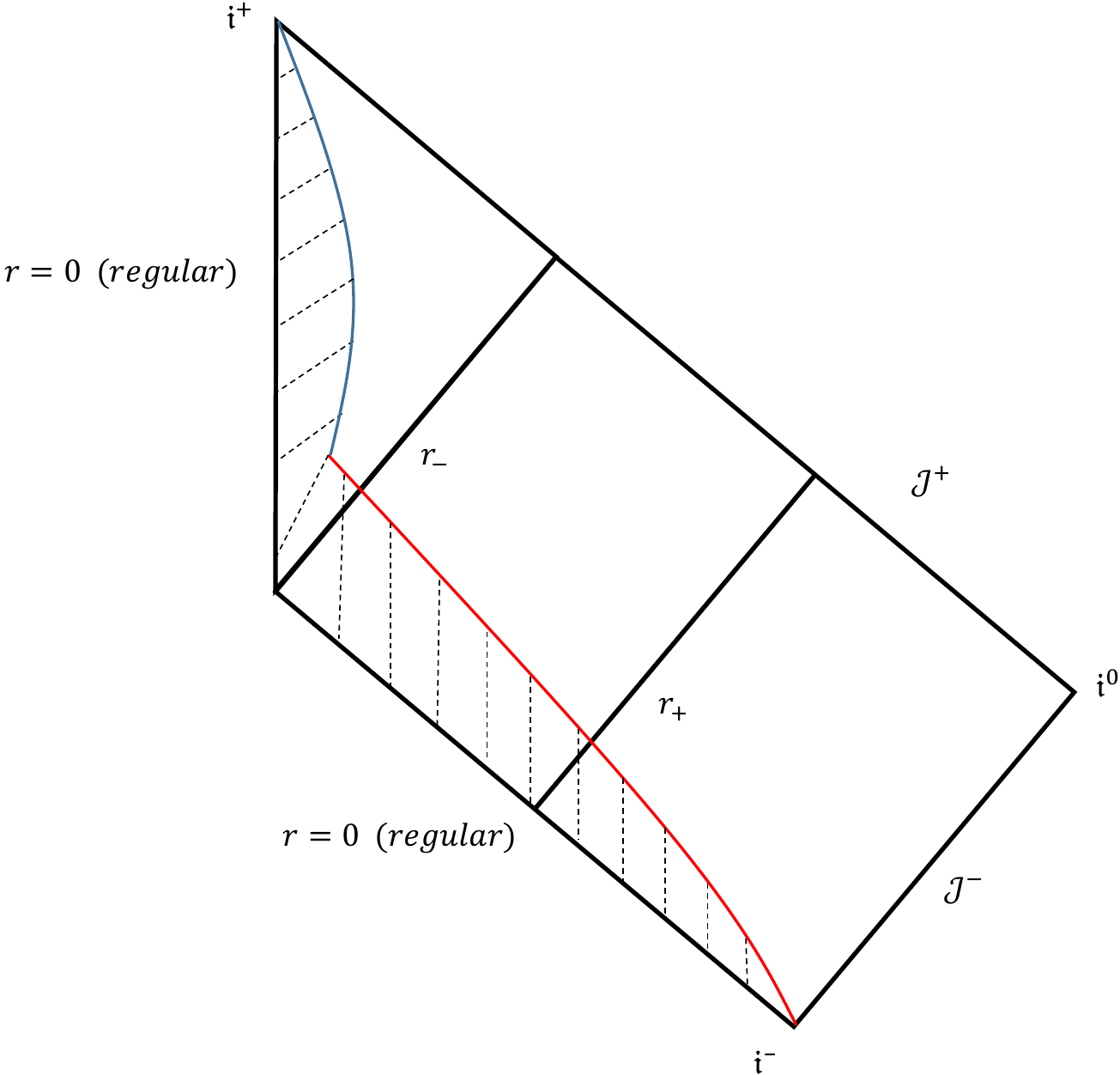}
\caption{\label{fig:Petroo}A simplified Penrose diagram for the homogeneous dust collapse with quantum vacuum effects. The red line represents the boundary of the collapsing object and the blue line shows the boundary surface of the ultimate stable static ball of dust inside the black hole. The areas with dashed lines represent the dust. The center of the black hole ($r=0$) is nonsingular.}
\end{figure}

This final fate of the collapse is a result of the existence of repulsive quantum vacuum forces which can be referred to as repulsive Casimir forces. These cause the matter to be trapped in a stable balanced situation inside the black hole. The important thing about this scenario is precluding the formation of the black hole singularity. To explain the reason, we remember that the physical radius of an arbitrary shell in the ball of dust with proper radius $r$ and proper time $\tau$ is defined as: $R(r,\tau )=a(\tau )r$ and therefore the singularity for any arbitrary shell of matter is reached when the scale factor goes to zero. Furthermore, the scale factor $a(\tau )$ is only a function of proper time and does not depend on the proper radius of the different shells of matter. Therefore in this scenario where the stable static ball of dust has the nonzero scale factor $a={{a}_{0}}=1$, the singularity does not form for the ball of dust and consequently for the corresponding black hole. In other words, a homogeneous ball of dust with a finite physical radius obviously does not admit a singularity. Thus the final fate of the homogeneous dust collapse with quantum vacuum effects, in certain conditions when $f(\xi)$ satisfies \eqref{eq:501}, will be a black hole wherein the trapped matter stays in a stable balanced situation inside the inner horizon, without any singularities. This outcome is an example of the resolution of singularity by employing quantum vacuum effects. Figure~\ref{fig:Petroo} shows the simplified Penrose diagram of the collapse.

\subsection{A physical example for the final fate of a dust star by considering quantum effects}
\label{sec:juju}

We Consider a spherically symmetric homogeneous dust star that has five times the mass of sun (${{M}_{m}}=5{{M}_{\odot }}$), with ${{M}_{\odot }}=1.989\times {{10}^{30}}kg$. Considering previous arguments, this object starts collapsing from an initial physical radius and continues it until passes the outer horizon and leads to the formation of a black hole. During this process, quantum vacuum energy density and pressure affect the interior and exterior geometries of the collapsing star. Ultimately, it will be stopped in a stable static phase inside the inner horizon of the black hole due to the repulsive quantum vacuum forces. In this example we investigate the physical features of the ultimate stable static point. As we argued before, we deal with negative values of $f(\xi )$ in \eqref{eq:41} and \eqref{eq:41prim} which correspond to the negative quantum vacuum energy densities and pressures which have repulsive effects. Furthermore, we obtained a condition for $f(\xi)$ in \eqref{eq:501} to attain real and positive values of ${\gamma }_{m}$ or equivalently Schwarzschild mass. To start calculation, we should choose a value for $f(\xi )$ that satisfies this condition. Here we assume that all couplings are equally possible. Considering figure~\ref{fig:coupli}, we can choose $f(\xi )=-0.002$ as typical value. Inserting this value into \eqref{eq:75}, for the static phase we get
\begin{equation}
\label{eq:79}
{{\gamma }_{m}}\simeq 310{{\gamma }_{r}}^{3}.
\end{equation}
Using \eqref{eq:49} we obtain
\begin{equation}
\label{eq:nnnwww}
{{\rho }_{0}}=\frac{18}{{{(8\pi )}^{2}}\times 0.002}{{\rho }_{p}}={{\rho }_{m}}=\frac{{{M}_{m}}}{\frac{4\pi }{3}{{R}_{c}}^{3}} \,
\end{equation}
where we have used the definition of the matter density for a homogeneous ball of dust. Using ${{M}_{m}}=5{{M}_{\odot }}$ and plugging the values of the Planck density and solar mass into \eqref{eq:nnnwww}, ${R}_{c}$ is calculated as
\begin{equation}
\label{eq:wownew}
{{R}_{c}}=1.97\times {{10}^{12}}{{\ell }_{p}} \,.
\end{equation}
Now with the help of \eqref{eq:72}, ${\gamma }_{r}$ is calculated. Inserting this parameter into \eqref{eq:79}, one can calculate ${{\gamma }_{m}}$ which, from \eqref{eq:72}, results in
\begin{equation}
\label{eq:81}
M=23.56\times {{10}^{38}}{{m}_{p}} \,.
\end{equation}
Note that as we discussed before, Schwarzschild mass $M$ is different from ordinary mass $M_{m}=0.914\times {{10}^{38}}{{m}_{p}}$ and the relative change is $\frac{\delta M}{M}=0.961$. Using ${\gamma }_{r}=1.97\times {{10}^{12}}$ and ${\gamma }_{m}=23.56\times {{10}^{38}}$ from above equations, the radii of the inner and outer horizons can be calculated by plugging these parameters into \eqref{eq:hior:76} and \eqref{eq:hior:77}. The result is
\begin{equation}
\label{eq:82}
{{r}_{+}} = 47.12\times {{10}^{38}}{{\ell }_{p}} \,,
\qquad
{{r}_{-}} = 2.03\times {{10}^{12}}{{\ell }_{p}} \,.
\end{equation}
Comparing the physical radius of the ball of dust from \eqref{eq:wownew} with the results in \eqref{eq:82} confirms our conclusion in the previous part that the physical radius in the stable static phase is always smaller than the inner horizon. In addition, the interior geometry of this static ball of dust which is a static closed FRW metric, according to \eqref{eq:48} and \eqref{eq:49}, has the curvature constant $k=59.68\frac{1}{{{\ell }_{p}}^{2}}$. These calculations can be repeated for any other masses of homogeneous balls of dust.

\subsection{A discussion on Hawking radiation}
\label{sec:hoho}

In the previous parts, we discussed that the ultimate fate of a collapsing homogeneous ball of dust with quantum vacuum effects of massless scalar fields with a specific range of couplings in the background. The final fate of the collapse was a black hole with an outer horizon which was the quantum corrected Schwarzschild horizon and a quantum originated inner horizon wherein the ball of dust stayed in a stable balanced situation. We argued that the existence of matter with a finite radius inside the inner horizon ensures that the black hole is devoid of the singularity. This means that the quantum vacuum effects are capable of resolving the singularity problem and one should not necessarily use quantum gravity for this purpose.

The natural question that arises is that whether Hawking radiation can change the previous predictions about the final fate of the collapse. The answer to this question is not straightforward since Hawking radiation itself is a first order quantum effect on a curved background. In fact, Hawking radiation results from the off-diagonal terms (traceless part) of the stress-energy tensor~\cite{dav,22}. Therefore, in principle, this phenomenon should be considered alongside the quantum vacuum effects that we applied in our model. However, simultaneous investigation of both effects makes the calculations formidable. Usually, in the case of Hawking radiation the background under consideration is regarded to be determined and one ignores any corrections due to the existence of Hawking flux on the background geometry itself. This is why we neglected the effects of Hawking flux on the background corrections in this work. In this regard, To consider the effects of Hawking radiation, we assume that the background geometry, which is corrected by quantum vacuum effects other than Hawking radiation, is determined and we study Hawking radiation in that.

After the onset of collapse, two different scenarios can be envisaged by taking Hawking radiation into account. Depending on the time scales of the collapse and evaporation, if the evaporation is faster, the collapsing ball of dust disappears before reaching the stable static phase. Then the black hole singularity will never form and one would not be worried about the singularity problem. The other scenario, in which the collapse is faster than the evaporation, allows the ball of dust to reach the stable static phase before the complete evaporation. In this case, after the ball of dust reaches the stable static point, with a radius that we calculated to be always smaller than the radius of the inner horizon, Hawking radiation continues shrinking the black hole and thus depleting the matter inside. The spherical symmetry of the background and distribution of Hawking flux points to the fact that the ball of dust will be diminished in an isotropic way so that its homogeneity and spherical symmetry will be preserved. Therefore Hawking radiation will shrink the radii of the horizons together with the physical radius of the ball of dust. Now we are ready to discuss the final steps of Hawking radiation in this simplified model.

The Hawking temperature of a black hole is defined by~\cite{hiak,hawk}
\begin{equation}
\label{eq:temp}
{T_{H}}=\frac{{\kappa }}{2\pi }
\end{equation}
where ${\kappa }$ is the surface gravity at the horizon. For a spherically symmetric black hole with the line element $d{{s}^{2}}=-f(r)d{{t}^{2}}+f{{(r)}^{-1}}d{{r}^{2}}+{{r}^{2}}d{{\Omega }^{2}}$, the temperature becomes
\begin{equation}
\label{eq:temhaw}
{{T}_{H}}=\frac{1}{4\pi }{{\left( \frac{df(r)}{dr} \right)}_{r={{r}_{+}}}}
\end{equation}
where ${r}_{+}$ is the outer horizon. In the case of our quantum corrected black hole, $f(r)$ is given by \eqref{eq:55}. Hawking radiation shrinks the radii of the inner and outer horizons \eqref{eq:hior:77} and \eqref{eq:hior:76} by decreasing ${\gamma }_{m}$ until they coincide (${r}_{+} = {r}_{-}$)\,. In this situation the Hawking temperature \eqref{eq:temhaw} vanishes and therefore the radiation stops. Note that this situation can be called "extremal" in analogy to Reissner-Nordstr\"{o}m black holes. The time scales of reaching this extremal phase would be much larger than the time needed for evaporation of classical Schwarzschild black holes. To see this we can compare the Hawking temperatures of quantum-corrected and classical black holes, with the same initial mass. Using the definition of ${\gamma }_{m}$ in \eqref{eq:72} and defining: ${{r}_{+}}=\alpha {{\ell }_{p}}$, from \eqref{eq:temhaw}, after working in ordinary units, we obtain
\begin{equation}
\label{eq:tonew}
{{T}_{H}}=\frac{{{m}_{p}}{{c}^{2}}}{2\pi {{k}_{B}}}\left( \frac{{{\gamma }_{m}}}{{\alpha}^{2}}-\frac{2}{45\pi }\frac{{{\gamma }_{m}}^{2}}{{\alpha}^{5}} \right)
\end{equation}
where ${{k}_{B}}$ is the Boltzmann constant and we have used \eqref{eq:55} for $f(r)$. For classical Schwarzschild black holes the temperature is given by
\begin{equation}
\label{eq:toneww}
{{T}_{H}}=\frac{{{m}_{p}}{{c}^{2}}}{2\pi {{k}_{B}}}\left( \frac{{{\gamma }_{m}}}{{\alpha}^{2}} \right) \,.
\end{equation}
From \eqref{eq:ohor}, we get
\begin{equation}
\label{eq:tonewww}
{\alpha}=2{{\gamma }_{m}}-\frac{1}{360\pi }\frac{1}{{{\gamma }_{m}}} \,,
\end{equation}
which by inserting it into \eqref{eq:tonew} and \eqref{eq:toneww}, we can find the Hawking temperatures of quantum-corrected and classical black holes as functions of ${\gamma }_{m}$.

\begin{figure}[tbp]
\centering
\includegraphics[scale=0.8]{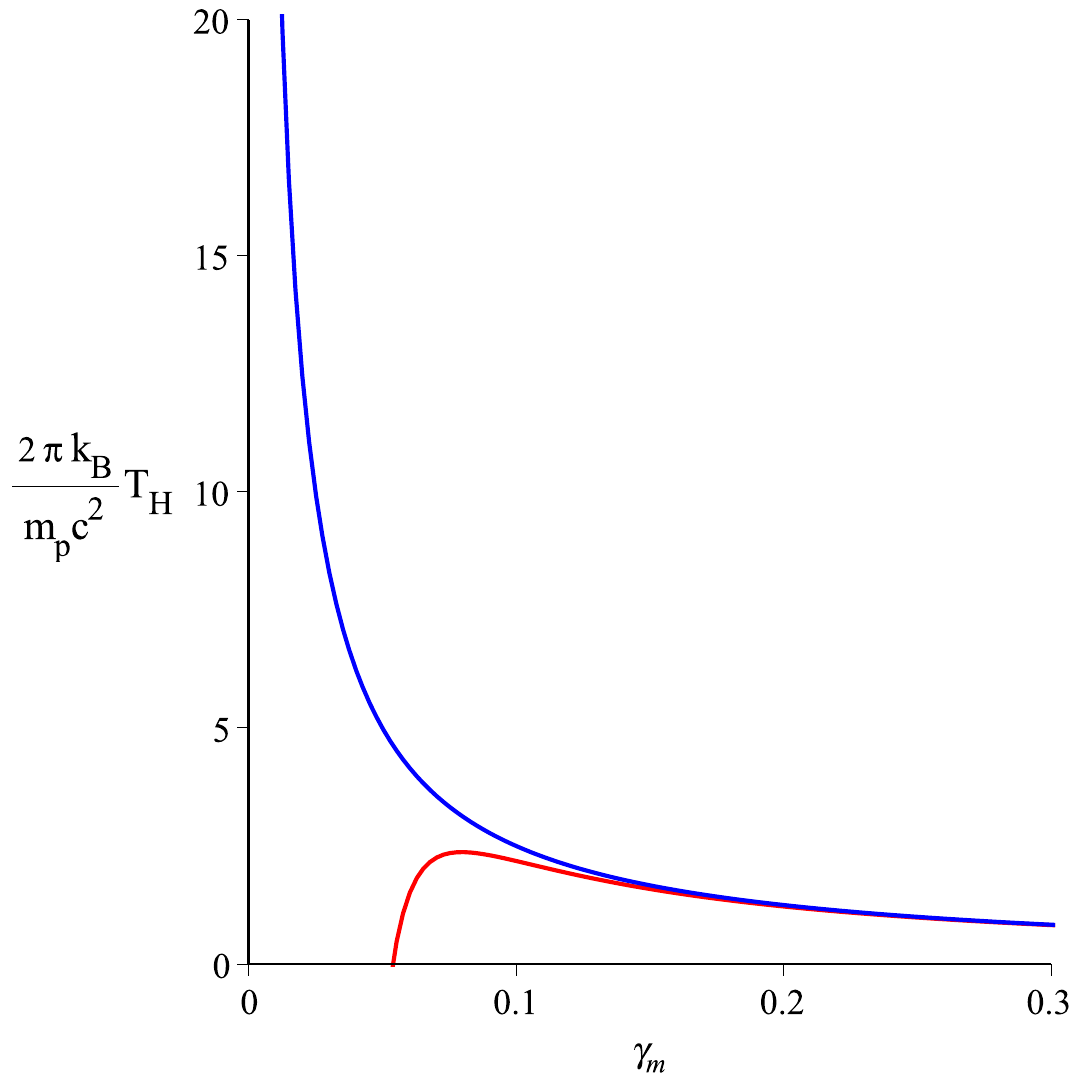}
\caption{\label{fig:aetem}A plot of normalized Hawking temperatures of classical (blue line) and quantum-corrected (red line) black holes as functions of ${\gamma}_{m}$. As the radiation continues, ${\gamma}_{m}$ reduces and the temperatures of both black holes increase in a similar way until ${{\gamma}_{m}}\approx 0.15$ which means $M\approx 0.15{m}_{p}$. From this point, quantum effects become extremely important and the temperatures start differing in behavior.}
\end{figure}

Figure~\ref{fig:aetem} represents the normalized Hawking temperatures of quantum corrected and classical black holes versus ${\gamma}_{m}$. As the Hawking radiation goes on, decreasing ${\gamma}_{m}$, the temperatures increase in an almost identical way. But as the Schwarzschild mass reaches about $0.15m_{p}$ (${{\gamma}_{m}}\approx 0.15$), differences start appearing. In this region the quantum effects become extremely important and cause the temperature of the quantum-corrected black hole to behave differently compared to the classical temperature. As we can see from figure~\ref{fig:aetem}, Hawking temperature of the quantum-corrected black hole starts decreasing from a specific value as the radiation continues, until it vanishes at the extremal point ${{\gamma}_{m}}\approx 0.054$. At this point the radiation stops. Note that the descending values of the quantum-corrected black hole temperature for values of ${{\gamma}_{m}}$ less than a specific value (about $0.07$) implies that the time scales of reaching the extremal phase will be very large. In fact the decrease of Hawking temperature from that point will decrease the mass loss rate, specially that of massive fields, and will prolong the process of reaching the extremal phase. This will be much more effective near the extremal point where the temperature is approaching zero, causing the rate of radiation to drop fast. In contrast, the classical black hole becomes hotter and hotter as its mass decreases, raising the mass loss rate due to the radiation, until it evaporates completely. Therefore the time it takes for our quantum-corrected black hole to reach the extremal phase will be longer than the evaporation time of the classical Schwarzschild black hole with the same initial mass.

The remarkable fact is the existence of a remnant of matter at the final extremal quantum-corrected black hole. This nonzero Schwarzschild mass implies a nonzero mass of matter(dust) which can again settle in a stable static situation and preclude the existence of singularity. The reason for this is that the right hand side of \eqref{eq:71}, which represents the contribution of ordinary mass (mass of dust) at the stable static point, will be nonzero for a nonzero Schwarzschild mass $M$. Therefore, even with Hawking radiation, the resolution of singularity will not be disrupted since a regular remnant will exist inside the extremal black hole.

To sum up, with all what was discussed in this section taken into account, regardless of which scenario is better for Hawking radiation of our quantum-corrected black hole, the resolution of singularity by quantum vacuum effects will not be negatively affected by the radiation.

\section{Final remarks}

In this article we showed that the expectation value of the quantum stress-energy tensor of massless scalar fields which are arbitrarily coupled to the background geometry, together with the energy density of a ball of homogeneous dust, can have important effects on the final fate of the collapse. We argued that for certain values of the coupling constants (that satisfy \eqref{eq:501}) the quantum background effects can be repulsive so that they can cancel out the attractive force of the gravitation of the dust and steer the collapsing system to a stable static phase after the formation of black hole. We showed that the static ball of dust always lies inside an inner horizon (with a quantum origin) and makes the black hole singularity-free. This happens since the interior geometry of the ball, described by a closed FRW metric, has nonzero scale factor. We also discussed that if Hawking radiation is considered, ultimately, the black hole becomes extremal with a remnant of matter (dust) inside. Again, this ultimate remnant will settle in a stable static phase and will preclude the formation of the singularity. Therefore Hawking radiation would not disrupt the resolution of the singularity.

As we calculated in section~\ref{sec:four}, at the moment the ball of dust enters the stable static region, the matter density is a factor of the Planck density. Furthermore, we showed that the maximum possible radius of the ball at the stable static phase will be ${{\left( \frac{1}{90\pi }\frac{M}{{{m}_{p}}} \right)}^{\frac{1}{3}}}{{\ell }_{p}}$. For ordinary black holes, this radius can be some orders of magnitude larger than the Planck length. In subsection~\ref{sec:juju} where a numerical example was brought for clarity, we calculated the matter (dust) density at the static stable phase to be ${{\rho }_{m}}\simeq 14.25{{\rho }_{p}}$ and the curvature constant to be $k=59.68\frac{1}{{{\ell }_{p}}^{2}}$\,. In addition, we calculated the radius of the ball of dust to be twelve orders of magnitude larger than the Planck length. Regarding the fact that the Planck densities and curvatures are the gates of the quantum gravity realm~\cite{pdff,17}, These values of the density and curvature in the stable static phase imply that on the cusp of quantum gravity the pure semi-classical effects would become extremely important. As a matter of fact, in this paper we have shown that the resolution of singularity, which is a task relegated to appropriate quantum gravity theories~\cite{gambini,corichi,vandersloot}, can also be attained by semi-classical methods like considering quantum vacuum effects on a curved background. In the following, we put a few remarks concerning some aspects and extensions of our considerations in order.

First, our work merely based on considering massless scalar fields. Taking the massless fields of arbitrary spin into account follows the same procedures that we implemented in this article except the coefficients will be spin-dependent. for example $f(\xi )$ in the quantum energy density of the interior region in \eqref{eq:31} will comprise other terms pertaining to different spins. In addition, the trace of the quantum-induced stress-energy tensor in \eqref{eq:fintr} will include other coefficients related to different values of spin. Although the structure of formulaes and calculations will be the same, these coefficients and their signs can be important in determining the final fate of the collapse.

Second, considering the discussions about Hawking radiation in section~\ref{sec:hoho}, the finite remnant of matter at the extremal situation implies that at least a finite amount of information can be preserved. This would alleviate the information loss paradox. Although some of the matter is radiated when the black hole reaches the extremality, it is interesting to investigate whether it is possible to store the information content in the final remnant of matter which lies inside the black hole. In remnant models of the information loss paradox~\cite{Giddings,Pchen} a similar final fate for the black hole is predicted. However, in these models the arguments about the final state of the remnant are usually based on quantum gravity theories. Here we have merely considered the quantum vacuum effects as a purely semi-classical method without regarding quantum gravity. This indicates that inclusion of quantum vacuum effects in collapse processes can lead to results that would not allow the information loss paradox to arise.

Finally, when we talk about the resolution of singularity or information loss paradox by quantum vacuum effects, as semi-classical predictions, some important questions arise about the genericity of such procedures. In fact one should investigate how much the final fate of the collapse would change if the physical configuration (which was a homogeneous dust in our example) changes by introducing inhomogeneities and/or ordinary (not quantum) pressures. It is important to realize that whether these changes can mathematically force the collapsing object to be crushed into zero volume, resulting in a singularity formation, or not. In this article we have assumed that the characteristic of dust as a pressureless material and the homogeneity condition remain unchanged during the collapse. It is noteworthy that the quantum-induced pressure and density will not disturb the homogeneous dust configuration since the existence and magnitude of these quantities are concomitants of this configuration. In other words, for different space-time configurations the elements of renormalized quantum vacuum stress-energy tensor will be different accordingly. That is because they have a geometric origin. For examining the genericity of the singularity resolution by quantum vacuum effects, one should calculate the quantum vacuum stress-energy tensors originating from more realistic physical configurations. These can be the interior and exterior geometries of a collapsing inhomogeneous dust, a collapsing perfect fluid or other collapsing objects including pressures and inhomogeneities. Then one can investigate whether the final outcome of the collapse would be devoid of the singularity or not. Such investigations will help us to understand what happens for the singularity resolution in our model if some ordinary pressures and/or inhomogeneities are involved perturbatively during the collapse. In this way the reliability of our model can be checked and we can understand whether our model can be generalized to more genuine collapse formalisms or not. If the answer is positive, we can establish semi-classical generic models for gravitational collapse that can resolve the singularity problem and even the information loss paradox. The answer to this question and the exact investigation of the previous two remarks can be interesting subjects to be studied elsewhere.

\acknowledgments

The authors are thankful to Carlos A. R. Herdeiro and Marco Sampaio for providing fruitful information in reference~\cite{herdeiro} where the figure~\ref{fig:coupli} is adopted from. We also thank Jahed Abedi for fruitful discussions and his constant interest.


\begin{thebibliography}{99}

\bibitem{a}
Pankaj S. Joshi, Daniele Malafarina, \emph{Recent developments in gravitational collapse and spacetime singularities}, \emph{Int. J. Mod. Phys. D}, {\bf20}, (2011) 2641  \href{https://arxiv.org/abs/1201.3660}{[arXiv:1201.3660]}.

\bibitem{b}
I.H. Dwivedi, P.S. Joshi, \emph{ Initial Data and the Final Fate of Inhomogeneous Dust Collapse}, \emph{Class. Quant. Grav.} {\bf14}, (1997) 1223--1236  \href{http://arxiv.org/abs/gr-qc/9612023}{[arXiv:gr-qc/9612023]}.

\bibitem{c}
Daniele Malafarina, Pankaj S. Joshi, \emph{ Gravitational collapse with non-vanishing tangential pressure}, \emph{Int. J. Mod. Phys. D}, {\bf20(4)}, (2011) 463 \href{http://arxiv.org/abs/1009.2169}{[arXiv:1009.2169]}.

\bibitem{d}
Seema Satin, Daniele Malafarina, Pankaj S. Joshi, \emph{Genericity aspects of black hole formation in the collapse of spherically symmetric slightly inhomogeneous perfect fluids}, \emph{ Int. J. Mod. Phys. D} {\bf25},(2016) 1650023 \href{https://arxiv.org/abs/1409.0505}{[arXiv:1409.0505]}.

\bibitem{6}
C. Bambi, D. Malafarina, and L. Modesto, \emph{Terminating black holes in asymptotically free quantum gravity}, \emph{Eur. Phys. J. C} {\bf74}, (2014) 2767 \href{http://arxiv.org/abs/1306.1668}{[arXiv:1306.1668]}.

\bibitem{7}
C. Bambi, D. Malafarina, and L. Modesto, \emph{Non-singular quantum-inspired gravitational collapse}, \emph{Phys. Rev. D} {\bf88}, (2013) 044009 \href{http://arxiv.org/abs/1305.4790}{[arXiv:1305.4790]}.

\bibitem{11}
L. Modesto, \emph{Disappearance of Black Hole Singularity in Quantum Gravity}, \emph{Phys. Rev. D} {\bf70}, (2004) 124009 \href{http://arxiv.org/abs/gr-qc/0407097}{[arXiv:gr-qc/0407097]}.

\bibitem{12}
R. Torres, \emph{Singularity-free gravitational collapse and asymptotic safety}, \emph{Phys. Lett. B} {\bf733}, (2014) 21--24 \href{http://arxiv.org/abs/1404.7655}{[arXiv:1404.7655]}.

\bibitem{13}
Ramon Torres, Francesc Fayos, \emph{On the quantum corrected gravitational collapse}, (2015) \href{http://arxiv.org/abs/1503.07407}{[arXiv:1503.07407]}.

\bibitem{16}
Cenalo Vaz, \emph{Quantum gravitational dust collapse does not result in a black hole}, \emph{Nuclear Physics B} {\bf891} (2015) 558--569 \href{http://arxiv.org/abs/1407.3823}{[arXiv:1407.3823]}.

\bibitem{18}
Sean A. Hayward, \emph{Formation and evaporation of non-singular black holes}, \emph{Phys. Rev. Lett.} {\bf96}, (2006) 031103 \href{http://arxiv.org/abs/gr-qc/0506126}{[arXiv:gr-qc/0506126]}.

\bibitem{radi}
Mandar Patil, Pankaj S. Joshi, Daniele Malafarina, \emph{Naked Singularities as Particle Accelerators II}, \emph{Phys. Rev. D} {\bf83}, (2011) 064007 \href{https://arxiv.org/abs/1102.2030}{[arXiv:1102.2030]}.

\bibitem{80}
P. S. Joshi and I. H. Dwivedi, \emph{Naked singularities in spherically symmetric inhomogeneous Tolman-Bondi dust cloud collapse}, \emph{Phys. Rev. D} {\bf47}, (1993) 5357. \href{http://arxiv.org/abs/gr-qc/9303037}{[arXiv:gr-qc/9303037]}.

\bibitem{bonanno}
Alfio Bonanno and Martin Reuter, \emph{Renormalization group improved black hole spacetimes}, \emph{Phys. Rev. D} {\bf62}, (2000) 043008. \href{https://arxiv.org/abs/hep-th/0002196}{[arXiv:hep-th/0002196]}.

\bibitem{Hossenfelder}
Sabine Hossenfelder, Leonardo Modesto, and Isabeau Prémont-Schwarz, \emph{Model for nonsingular black hole collapse and evaporation}, \emph{Phys. Rev. D} {\bf81}, (2010) 044036. \href{https://arxiv.org/abs/0912.1823}{[arXiv:0912.1823[gr-qc]}

\bibitem{casadio}
Roberto Casadio, \emph{On gravitational fluctuations and the semiclassical limit in minisuperspace models}, \emph{Int. J. Mod. Phys. D} {\bf09}, 511 (2000). \href{https://arxiv.org/abs/gr-qc/9810073v3}{[arXiv:gr-qc/9810073]}.

\bibitem{barcelo}
Carlos Barcel\'{o}, Stefano Liberati, Sebastiano Sonego, and Matt Visser, \emph{Fate of gravitational collapse in semiclassical gravity}, \emph{Phys. Rev. D }{\bf77}, (2008) 044032. \href{https://arxiv.org/abs/0712.1130}{[arXiv:0712.1130 [gr-qc]]}.

\bibitem{jah.arf}
Jahed Abedi, Hessamaddin Arfaei, \emph{Obstruction of black hole singularity by quantum field theory effects}, \emph{J. High Energ. Phys.}(2016) 2016: 135 \href{http://arxiv.org/abs/1506.05844}{[arXiv:1506.05844 [gr-qc]]}.

\bibitem{dav}
N. D. Birrell and P. C. W. Davies, \emph{Quantum fields in curved space}, Cambridge University Press (1982).

\bibitem{parker}
Leonard E. Parker, David J. Toms, \emph{Quantum field theory in curved spacetime}, Cambridge University Press (2009).

\bibitem{bunch}
T.S. Bunch and L. Parker, \emph{Feynman propagator in curved spacetime: A momentum-space representation}, \emph{Phys. Rev. D} {\bf20}, (1979) 2499.

\bibitem{j}
Bryce S. DeWitt, \emph{Quantum field theory in curved spacetime}, \emph{Phys. Rep.} {\bf19}, no.6 (1975) 295--357.

\bibitem{k}
Julian Schwinger, \emph{On Gauge Invariance and Vacuum Polarization}, \emph{Phys. Rev.} {\bf82}, (1951) 664.

\bibitem{n}
J. H. Van Vleck, \emph{The Correspondence Principle in the Statistical Interpretation of Quantum Mechanics}, \emph{Proc.Nat.Acad.Sci.} {\bf(USA),14} (1928) 178.

\bibitem{schwarzqft}
Matthew D. Schwartz, \emph{Quantum field theory and the standard model}, Cambridge University Press, March(2014).

\bibitem{OS}
J. R. Oppenheimer and H. Snyder, \emph{On continued gravitational contraction}, \emph{Phys Rev.} {\bf56}, (1939) 455.

\bibitem{Datt}
S. Datt, \emph{Zs.f.Phys.} {\bf108} (1938) 314.

\bibitem{radial}
R. Cai, L. Cao, N. Ohta, \emph{Black Holes in Gravity with Conformal Anomaly and Logarithmic Term in Black Hole Entropy}, \emph{J. High Energ. Phys.}(2010) 2010: 82 \href{http://arxiv.org/abs/0911.4379}{[arXiv:0911.4379]}.

\bibitem{ford1976}
L. H. Ford, \emph{Quantum vacuum energy in a closed universe}, \emph{Phys. Rev. D} {\bf14}, (1976) 3304.

\bibitem{herdeiro}
Carlos A. R. Herdeiro and Marco Sampaio, \emph{Casimir energy and a cosmological bounce} \emph{Class. Quant. Grav.} {\bf23}, (2006) 473--484 \href{https://arxiv.org/abs/hep-th/0510052}{[arXiv:hep-th/0510052]}.

\bibitem{dowker1977}
J. S. Dowker and Raymond Critchley, \emph{Vacuum stress tensor in an Einstein universe: Finite-temperature effects}, \emph{Phys. Rev. D} {\bf15}, (1977) 1484.

\bibitem{poisson}
Eric Poisson, \emph{A relativist's toolkit: The mathematics of black-hole mechanics}, Cambridge University Press (2004).

\bibitem{22}
W. G. Unruh, \emph{Notes on black-hole evaporation}, \emph{Phys. Rev. D} {\bf14}, (1976) 870.

\bibitem{hiak}
S. W. Hawking, \emph{Black hole explosions?}, \emph{Nature} {\bf248}, (1974), 30-31.

\bibitem{hawk}
S. W. Hawking, \emph{Black holes and thermodynamics}, \emph{Phys. Rev. D} {\bf13}, (1976), 191.

\bibitem{Giddings}
Steven B. Giddings, \emph{Black holes and massive remnants}, \emph{Phys. Rev. D} {\bf46}, (1992)  1347 \href{https://arxiv.org/abs/hep-th/9203059}{ 	 [arXiv:hep-th/9203059]}.

\bibitem{Pchen}
P. Chen,  Y.C. Ong,  D.-h. Yeom, \emph{Black hole remnants and the information loss paradox}, \emph{Physics Reports}, Vol {\bf603} (2015) 1–45 \href{https://arxiv.org/abs/1412.8366}{[arXiv:1412.8366 [gr-qc]]}

\bibitem{pdff}
Hal M. Haggard and Carlo Rovelli, \emph{Quantum-gravity effects outside the horizon spark black to white hole tunneling}, \emph{Phys. Rev. D} {\bf92}, (2015) 104020 \href{https://arxiv.org/abs/1407.0989}{[arXiv:1407.0989]}.

\bibitem{17}
Abhay Ashtekar, Tomasz Pawlowski, Parampreet Singh, and Kevin Vandersloot, \emph{Loop quantum cosmology of k=1 FRW models}, \emph{Phys. Rev. D} {\bf75}, (2007) 024035 \href{https://arxiv.org/abs/gr-qc/0612104}{[arXiv:gr-qc/0612104]}.

\bibitem{gambini}
Rodolfo Gambini and Jorge Pullin, \emph{Loop quantization of the schwarzschild black hole}, \emph{Phys. Rev. Lett.} {\bf110}, (2013) 211301 \href{https://arxiv.org/abs/1302.5265v2}{[arXiv:1302.5265]}.

\bibitem{corichi}
Alejandro Corichi and Parampreet Singh, \emph{Loop quantization of the Schwarzschild interior revisited}, \emph{Class. Quantum Grav.} {\bf33}, (2016) 055006 \href{http://arxiv.org/abs/1506.08015}{[arXiv:1506.08015]}.

\bibitem{vandersloot}
Christian G. Bohmer and Kevin Vandersloot, \emph{Loop quantum dynamics of the Schwarzschild interior}, \emph{Phys. Rev. D} {\bf76}, (2007) 104030 \href{http://arxiv.org/abs/0709.2129}{[arXiv:0709.2129]}.

\end{thebibliography}
\end{document}